\documentclass{article}
\usepackage{amssymb}
\usepackage{amsmath}
\usepackage{a4wide}
\usepackage{latexsym}
\usepackage{epsfig}
\begin{document}
\newcommand{\fr}{{\cal F}}
\newcommand{\systems}{{\cal H}}
\newcommand{\g}{{\cal G}}
\let\bigand=\bigwedge
\let\iso=\cong
\let\eq=\cong
\let\b=\Box
\let\projected=\uparrow
\let\implies=\Rightarrow

\newtheorem{xproof}{Proof}
\renewcommand{\thexproof}{}  %
\newif\ifSuppressqed\Suppressqedfalse
\newenvironment{proof}{\global\Suppressqedfalse\begin{xproof}}{
        \ifSuppressqed\global\Suppressqedfalse
        \else\xqed\fi\end{xproof}}
\newcommand{\xqed}{\pushright{$\Box$}}
\newcommand{\qed}{\xqed\global\Suppressqedtrue}
\let\la=\langle
\let\ra=\rangle
\let\proofnq\proof
\let\endproofnq\endproof

\let\biimplies=\Leftrightarrow
\let\bo=\bigotimes
\let\union=\bigcup
\let\bigunion=\bigcup
\let\intersection=\bigcap
\let\proves=\vdash
\let\to=\rightarrow
\let\sat=\models
\let\false=\perp
\let\fusion=\bo
\let\bigand=\bigwedge
\let\bigor=\bigvee
\let\next=\bigcirc
\let\cons=\vdash
\let\union=\bigcup
\let\composed=\circ
\let\dia=\Diamond
\let\d=\Diamond
\def\pushright#1{{\parfillskip=0pt\widowpenalty=10000
  \displaywidowpenalty=10000\finalhyphendemerits=0
  \leavevmode\unskip\nobreak\hfil\penalty50\hskip.2em\null
  \hfill{#1}\par}}
\def\S{{\ifmmode\mathord{{\fam0 S5WD}_n}\else S5WD$_n$\fi}}
\def\Ss{{\ifmmode\mathord{{\fam0 S5WD}^*_n}\else S5WD^*$_n$\fi}}
\def\Sss{{\ifmmode\mathord{{\fam0 S5WD}^{**}_n}\else S5WD^{**}$_n$\fi}}
\newcommand{\cC}{{\cal C}}
\newcommand{\bfut}[0]{G}
\newcommand{\bpast}[0]{H}
\newcommand{\dfut}[0]{F}
\newcommand{\dpast}[0]{P}
\newcommand{\since}[2]{#1 {\cal S} #2}
\newcommand{\until}[2]{#1 {\cal U} #2}
\newcommand{\sinceuno}[2]{#1 {\cal S}_{1} #2}
\newcommand{\untiluno}[2]{#1 {\cal U}_{1} #2}
\newcommand{\sincedue}[2]{#1 {\cal S}_{2} #2}
\newcommand{\untildue}[2]{#1 {\cal U}_{2} #2}
\newcommand{\lang}[1]{{{\cal L}_{#1}}}
\newcommand{\model}[1]{{\cal M}_{#1}}
\newtheorem{definition}{Definition}[section]
\newtheorem{theorem}{Theorem}[section]
\newtheorem{lemma}{Lemma}[section]
\newtheorem{corollary}{Corollary}[section]
\newtheorem{observation}{Observation}[section]

\newtheorem{example}{Example}[section]

\newcommand{\vp}{\varphi} 
\newcommand{\cL}{{\cal L}} 
\newcommand{\cF}{{\cal F}}

\title{Knowledge in Multi-Agent Systems:\\
Initial Configurations and Broadcast} 

\author{
Alessio Lomuscio\\
Dept of Electronic Engineering\\
QMW College, University of London\\
London, UK \and 
Ron van der Meyden\\
Computing Sciences\\
University of Technology\\ Sydney, Australia
\and 
Mark Ryan\\
School of Computer Science\\
University of Birmingham\\
Birmingham, UK
}

\maketitle

\bibliographystyle{alpha}
\begin{abstract}
  The semantic framework for the modal logic of knowledge due to
  Halpern and Moses provides a way to ascribe knowledge to agents in
  distributed and multi-agent systems.  In this paper we study two
  special cases of this framework: \emph{full systems} and
  \emph{hypercubes}.  Both model static situations in which no agent
  has any information about another agent's state.  Full systems and
  hypercubes are an appropriate model for the initial configurations
  of many systems of interest.  We establish a correspondence between
  full systems and hypercube systems and certain classes of Kripke
  frames.  We show that these classes of systems correspond to the
  same logic. Moreover, this logic is also the same as that generated
  by the larger class of {\em weakly directed frames}.  We provide a
  sound and complete axiomatization, \S, of this logic.
Finally, we show that under certain natural assumptions, in a model
where knowledge evolves over time, \S{} characterizes the
properties of knowledge not just at the initial configuration, but
also at all later configurations. In particular, this holds for
\emph{homogeneous broadcast systems}, which capture settings in which
agents are initially ignorant of each others local states, operate
synchronously, have perfect recall and can communicate only by
broadcasting.
\end{abstract}

\newpage
 
\section{Introduction}

{\em Modal logics of knowledge} have been proposed as a formal tool
for specifying and reasoning about multi-agent systems in a number of
disciplines, including Distributed Computing \cite{HalpernMoses90},
Artificial Intelligence \cite{jmc:ai:logic} and Economics
\cite{Aumann76,RubensteinWolinsky90}.

The logic most commonly applied in this area is the logic S5$_n$, a
generalization to a multi-agent setting of the logic S5 (see, e.g.,
\cite{HughesCresswell96} and
\cite{Goldblatt}), which was
originally proposed as a model of knowledge by Hintikka
\cite{jh:kb:cp} in Philosophical Logic.
This application of S5 interprets the modal formula $\b \phi$ 
as ``it is known that $\phi$''. So interpreted, 
the logic S5  models an \emph{ideal} agent, whose knowledge has
the properties of being veridical (everything the agent knows is
true), and being closed under positive introspection (the agent knows
what it knows) and negative introspection (it knows what it does
not know). These properties have been the subject of a significant
body of criticism in the philosophical literature, but S5 remains an
appropriate model for applications in computer science and economics,
since it captures an information theoretic notion of knowledge of
interest in these areas.

The logic S5$_n$ is a multi-modal version of S5, including for
each $i=1\ldots n$ an operator $\b_i$.  The intended interpretation is
that each $i=1,\dots, n$ represents an agent, and $\b_i\phi$ expresses
``agent $i$ knows that $\phi$.'' The logic S5$_n$ can be axiomatized by taking
all the propositional tautologies; the axiom schemas 
$ \b_i(p
\implies q) \implies \b_i p \implies
\b_i q$, and $ \b_i p \implies p$, and $\b_i p \implies \b_i \b_i
p$, and $\neg \b_i \neg p \implies \b_i \neg \b_i \neg p$, 
and the inference rules Modus Ponens, Necessitation and Uniform
Substitution.

The logic S5$_n$ has also been extended to deal with properties that
arise when we investigate the state of knowledge of the group.  Subtle
concepts like common knowledge and distributed knowledge have been
investigated, as has the combination of the logic of knowledge and time
(see \cite{fhmv:rak,MeyerHoek} for extensive treatments of this 
literature.) 
Although the focus of research in this area has been on the
combination of the knowledge modalities, modeled by S5, with
modalities expressing other mental states, it has been noted that in
certain situations, the axioms of \S{} provide an incomplete
description of the properties of knowledge.  One of the
characteristics of the S5$_n$ axioms is that they do not appear to
state any interaction between one agent's knowledge and that of
another. Some such interactions nevertheless follow, e.g., it can be
shown that $\b_i \phi \implies \neg \b_j \neg \phi$ is valid in
S5$_n$.  (This formula above states that agent $j$ cannot rule out the
possibility of a fact known by agent $i$.)  However, there are
specific settings in which further such interactions hold.  For
example, consider a distributed system composed of a group of agents
$A=\{1,\dots,n\}$ and the
following situations:
\begin{description}
\item[One agent knowing everything the others know.] An agent $j$
is the 
central librarian 
of a distributed system of agents
that  rely on $j$ to 
maintain all their knowledge.
\item[Linear order in agents' private knowledge.] 
The agents operate within a chain of command subject to security restrictions. 
Each agent in the chain has a higher security clearance 
than the previous agent, and has access to a larger set of information
sources. 
\end{description}
These and similar scenarios can 
be modeled by
extensions of S5$_n$ in which \emph{interaction axioms} are imposed.
Write $S_{i,j}$ for the axiom schema 
$\b_i \phi \implies \b_j \phi$. 
Then the  first example 
above can be modeled by the
logic S5$_n$ plus $S_{i,j}$ for all $i\in A$. 
The second scenario can be described by assuming an order on the set
of agents 
reflecting their increasing
information, and by taking S5$_n$ plus $S_{i,j}$ 
for all $i\leq j$. 
These are just two isolated examples but there is actually a
\emph{broad spectrum} of possible specifications on how private states
of knowledge are affected by other agents' knowledge
(see \cite{LomuscioRyan99} for a detailed exposition).
At one end
of the spectrum we have the system S5 in which 
all 
agents have the same knowledge. This can be modeled by taking an
extension of S5$_n$ in which the axiom
$S_{i,j}$ holds for \emph{all} $i,j\in A$, 
making all the modalities collapse onto each other. This is a
very strong constraint.  At the other end of the spectrum is simply
S5$_n$. 
Catach 
\cite{Catach88} has studied a
limited class of such interactions between 
knowledge of the agents.

While the examples of interaction axioms above derive directly from
static assumptions about the interaction between agents' knowledge,
there are also cases where such interactions arise in more subtle
ways. Rather than start with assumptions that are directly about
interactions between agents' knowledge, one could begin with an
\emph{extensional} model of distributed systems, of the kind commonly
used in studies of distributed computing.  The notion of
\emph{interpreted system} of Halpern and Moses
\cite{HalpernMoses90,fhmv:rak} takes this approach.  The interpreted systems
model describes a multi-agent system in terms of the \emph{local
states} of the agents and how such local states evolve over time as
the agents communicate. A local state may be as mundane as a listing
of the values of the set of variables maintained by the agent, or it
could be a richly structured representation of the information available
to the agent. A \emph{global state} consists of a local state for each agent,
plus a state for the environment within which the agents operate.  In
general, there may be constraints 
connecting
the components of a global
state, so not every possible global state need occur in the system.

One can define two situations in such a system to be equivalent for
agent $i$ if the agent has the same local state in these situations.
This equivalence relation can then be used as the accessibility
relation corresponding to 
agent $i$'s knowledge operator $\b_i$.  This
approach provides an information theoretic notion of knowledge that
has been been found useful in analyses of distributed systems (see
\cite{fhmv:rak} for discussion of a number of examples and extensive
citations.)

Generally, the logic of knowledge that arises from this 
semantic framework is S5$_n$. However, it has been noted that
in certain quite natural special cases, additional 
axioms arise that state interactions between agents knowledge. 
Fagin et al.\ \cite{JACM::FaginHV1992,FaginVardi86}
present one such example. They study systems in which 
agents with perfect recall, operating in a static world, 
communicate their knowledge about that world by means of unreliable 
message passing.

In this paper we introduce and study another special 
case of the interpreted systems model that results in 
interaction axioms 
and can be axiomatized by an extension of S5$_n$ that falls into
the above-mentioned 
spectrum. The classes of systems we investigate are called \emph{full
systems} and \emph{hypercubes}. Both are systems in which \emph{every}
possible combination of individual agents' local states occurs in some
global state in the system. In hypercubes we require additionally that
every combination of state of the environment and the agents' local
states occurs.
Full systems and hypercubes are appropriate classes of systems for
modeling the initial configurations of many systems of interest, in
which no agent has any information concerning any other agent's state.
(Thus each agent considers possible every combination of the other
agents' local states.)

Full systems and hypercubes may be shown to satisfy an axiom that does
not follow from S5$_n$. This axiom states in a quite intuitive fashion
the property that every combination of the individual  agent's local states 
occurs in some global state of the system.  By characterizing full
systems and hypercubes in terms of certain classes of Kripke frames,
we establish a sound and complete axiomatization of the logic of
knowledge in these classes of systems.  Interestingly, the two classes
correspond to the \emph{same} logic, which we call \S.  The
nomenclature arises from the fact that we show that a further class of
frames, the \emph{weakly directed} frames, corresponds to the same
logic.  We also show that \S{} is decidable.

The definition of full systems and hypercubes takes a static viewpoint
of multi-agent systems that does not use the full power of the
interpreted systems model, which is also capable of modeling the
evolution of knowledge over time.  As noted above, these definitions
provide an appropriate characterization of the agents' knowledge in the
initial configurations of many distributed systems.  However, we 
show that the logic \S{} has broader applicability than simply
reasoning about such initial configurations.  We also study in this
paper the dynamic behavior of knowledge in
\emph{homogeneous broadcast environments}. These model 
a particular communication architecture, in which agents operate
synchronously and can communicate only by broadcasting information to
\emph{all} agents.  We assume that agents have perfect recall, and
that their initial configuration is characterized by a hypercube.  We
show that not just the initial configuration, but \emph{all}
configurations arising in such a system can be characterized using a full
system.  It follows that \S{} exactly captures  the properties of
knowledge in homogeneous broadcast systems.  Since 
\S{} extends S5$_n$, this provides another example of a natural
situation in which S5$_n$ is an incomplete characterization of the
logic of knowledge, analogous to the results of Fagin et al.\
\cite{JACM::FaginHV1992,FaginVardi86}.

The paper is organized as follows. In Section~2 we recall the two
standard semantics for knowledge in multi-agent systems (Kripke models
and interpreted systems), and we introduce full systems and
hypercubes. In Section~3 we formally relate full systems and
hypercubes to Kripke models by identifying corresponding classes of
Kripke frames. We also show that with respect to the logic of
knowledge we consider in this paper these classes of frames generate
the same logic. In Section~4 we present a sound and complete
axiomatization \S{} for this logic. We prove the logic decidable
in Section~5. These sections all deal with a static framework. In
Section~6 we go on to consider a dynamic framework that models how
agents' knowledge changes over time.  We define homogeneous broadcast
systems, and show that agents' states of knowledge in such systems can
be characterized by a hypercube at each point of time, thereby showing
that \S{} is also a sound and complete axiomatization of the logic
of knowledge in homogeneous broadcast systems. 
We illustrate the theory with an example of a two-person card game. 
Finally, in Section~7 we draw our conclusions and we suggest further
work.

\section{Definitions}

Amongst the approaches that have been proposed to the semantics of 
logics for knowledge are \emph{interpreted systems} and \emph{Kripke models}.
The two approaches have different advantages and disadvantages. On the
one hand, interpreted systems provide a more concrete and intuitive
way to model real systems, but on the other hand Kripke models come
with an heritage of fundamental techniques that may be used to prove
properties of the logic.

In this section we briefly recall the key definitions of Kripke frames
and interpreted systems.  We then we define hypercube systems and full
systems, the particular classes of systems that are the focus of this
paper.

We use the following mathematical notations throughout. 
If $W$ is a set, we write $|W|$ for its cardinality. 
If $\sim$ is an equivalence relation on $W$ and $w\in W$, then we
write $W/{\sim}$ for the set of equivalence classes of $\sim$, and
write $[w]_{\sim}$ for the equivalence class containing $w$.

\subsection{Kripke models}

Kripke models \cite{Kripke} were first formally proposed in
Philosophical Logic. %
They have since been used within 
computer science and Artificial Intelligence 
as semantic
structures for logics for belief, logics for knowledge, temporal
logics, logics for actions, etc., all of which are modal logics.  Over
the last thirty years, many formal techniques have been developed for
the study of modal logics grounded on Kripke semantics, such as
completeness proofs via canonical models, decidability via the finite
model property \cite{HughesCresswell96}, and more recently, techniques
for combining logics \cite{KrachtWolter:jsl,Gabbay:Fibring1}.

We now briefly recall a few concepts from this literature that we will
be using later in the paper. 
For more technical details and
motivation, the reader is referred to an 
introduction to modal
logic, such as \cite{HughesCresswell96} or \cite{Goldblatt} or
\cite{Hughes+Cresswell:1984}. 
We state our definitions for the multi-modal case, which is a slight
generalization of those in much of the literature.

\newcommand\atoms{{\it Atoms}}

We assume a set 
$\atoms =\{p,\dots\}$ 
of \emph{propositional atoms}, and a
finite $A=\{1,\dots,n\}$ of \emph{agents}.  We will deal primarily with 
a formal language 
given by the following grammar:
$$\phi\;::=\; p\mid \neg\phi \mid \phi_1\land \phi_2\mid \b_i \phi$$ where 
$p\in \atoms$ and $ i\in A$. 
We write $\cL_n$ for the set of formulae generated by this grammar when 
$A=\{1,\dots,n\}$. 
Intuitively the formula $\b_i
\phi$ represents the situation in which the agent $i$ knows the fact
represented by the formula $\phi$. Other propositional connectives
such as disjunction and implication 
can be defined in the usual way.   
If $W$ is a set, then $id_W$ is the identity relation on $W$, i.e.,
the relation $\{(w,w) \mid w\in W\}$.

Kripke semantics is based on the following structures.

\begin{definition}[Kripke frames and Kripke models]\label{frames}
A \emph{frame} $F$ is a tuple $F=(W,R_1, \dots,\linebreak R_n)$, where $W$ is a
  non-empty set 
(called set of worlds)
  and 
  for each $i \in A$ the component $R_i$
  is a binary
  relation on $W$. If all relations are equivalence
  relations, the frame is an \emph{equivalence frame} and we write
  $\sim_i$ for $R_i$.

A \emph{model} $M$ is a tuple $M=(W,R_1, \dots, R_n, \pi)$,
  where $(W,R_1, \dots, R_n)$ is 
  a frame called 
  its underlying frame and 
$\pi: W \to 2^{\atoms}$ is an interpretation for the atoms.  
An \emph{equivalence
    model} is a model whose underlying frame is an equivalence frame.
\end{definition}
We will call equivalence frames {\em E frames} and equivalence models
{\em E-models}.  The class of equivalence frames will be denoted by
${\cal F}_E$.
The class of all equivalence models is frequently taken to be the
appropriate class of structures for the logic of knowledge.\footnote {
Philosophers have long held qualms about properties of knowledge (such
as negative introspection) that are consequences of this class of
structures \cite{Lenzen78}.  For computer scientists and economists,
however, equivalence frames capture an information theoretic notion of
knowledge that is useful for their applications. } The logic
corresponding to this class of structures is the logic S5$_n$
\cite{HalpernMoses90}.

Committing an abuse of notation, given a frame $F=(W,R_1, \dots,
R_n)$, and an interpretation $\pi$, we will sometimes denote
$M=(W,R_1, \dots, R_n, \pi)$ as $M=(F,\pi)$. Also when the set
$A=\{1,\dots, n\}$ is clear from the context, we will denote
$M=(W,R_1, \dots, R_n, \pi)$ as $M=(W, \{R_i\}_{i \in A},\pi)$ and
$F=(W,R_1, \dots, R_n)$ as $F=(W,\{R_i\}_{i \in A})$.

\begin{definition}[Satisfaction]\label{satisfaction}
  The satisfaction of a formula $\phi$ in a world $w$ of a model
  $M$, formally $M\sat_w \phi$ is inductively defined as follows:
\begin{tabbing}
  $M\sat_w \phi\land \psi$ X XXXXX\=\kill
  \>  $M\sat_w p$           \' if  $p\in \pi(w)$ \\
  \>  $M\sat_w \neg \phi$      \' if  $M\not\sat_w \phi$\\
  \>  $M\sat_w \phi\land \psi$    \' if $M\sat_w \phi \mbox{ and } M\sat_w \psi$\\
\>  $M\sat_w \b_i \psi$    \' if for each $w'\in W$, $w R_iw'$
implies $M\sat_{w'} \psi$\\
\end{tabbing}
Satisfaction for the other logical connectives can be defined in the
usual way.
\end{definition}

Validity is also defined by means of the standard definition:
\begin{definition}[Validity]\label{val-frames}
  A formula $\phi$ is valid on a model $M=(W,R_1, \dots, R_n,\pi)$, if
  for 
  every point $w \in W$ we have $M\sat_w \phi$. A formula $\phi$ is
  valid on a frame $F=(W,R_1, \dots, R_n)$ if for every interpretation
  $\pi$ we have $(F,\pi)\sat \phi$. A class of models $\cal{M}$
  validates a formula $\phi$, 
  denoted $\cal{M} \models \phi$, 
 if for every model $M\in \cal{M}$ we have
  $M\sat \phi$. A formula $\phi$ is valid on a class of frames
  $\cal{F}$, 
  denoted $\cal{F} \models \phi$, 
 if for every frame $F\in \cal{F}$ we have $F\sat \phi$.
\end{definition}

Two frames $F = (W, R_1, \ldots, R_n)$ and $F' = ( W', R_1',
\ldots, R_n')$ are said to be {\em isomorphic} if there exists a
bijective function $h: W\rightarrow W'$ such that for each $i=1\dots
n$ and all points $w,w'\in W$ we have $w R_i w'$ if and only if $h(w) R_i' h(w')$. 
We write $F \equiv F'$ when this is the case. 

We clearly have the following.
\begin{theorem} \label{thm:fiso} 
If $F$ and $F'$ are frames with $F \equiv F'$ then for all $\psi\in
\cL_n$ we have $F\models \psi$ if and only if $F'\models \psi$.
\end{theorem}

Slightly more general than the notion of frame isomorphism is the
notion of {\em p-morphism}. We can define these both at the level of
frames and at the level of Kripke models.

\begin{definition} [p-morphism]
Let $F = (W, R_1, \ldots, R_n)$ and $F' = (W', R_1', \ldots, R_n')$
be frames. A {\em frame p-morphism} from $F$ to $F'$ is a mapping 
$p:W\rightarrow W'$ that satisfies 
\begin{enumerate} 
\item the function $p$ is surjective, and 
\item for all $u,v\in W$ and each $i=1\ldots n$, if 
      $uR_i v$ then $p(u)R_i p(v)$, and 
\item for each $i=1\ldots n$ and $u\in W$ and $v'\in W'$, if $p(u) R_i' v'$ 
    then there exists $v\in W$ such that $u R_i v$ and $p(v) = v'$.
\end{enumerate} 
If $M = (W, R_1, \ldots, R_n, \pi)$ and $M' = (W', R_1', \ldots, R_n',
\pi')$ are Kripke structures, then a {\em model p-morphism} from $M$
to $M'$ is a mapping $p:W\rightarrow W'$ that is a frame p-morphism
from $(W, R_1, \ldots, R_n)$ to $(W', R_1', \ldots, R_n')$
and satisfies $q \in \pi'(p(w))$ if and only if $q \in \pi(w)$ for all
propositions $q \in \atoms$ and all worlds $w \in W$.
\end{definition}

The following result shows that $p$-morphisms preserve 
satisfaction and validity for the language $\cL_n$. 

\begin{theorem} \label{thm:pmorph1} 
(\cite{Hughes+Cresswell:1984} page 73)
If $p$ is a model p-morphism from $M$ to $M'$ then 
for all worlds $w$ of $M$ and formulae $\vp \in \cL_n$, we have
$M\models_w \vp $ if and only if $M'\models_{p(w)}\vp $. Thus $\vp$ is valid on
$M$ if and only if $\vp$ is valid on $M'$.  
If $p$ is a frame p-morphism from $F$ to $F'$ then for all $\vp \in
\cL_n$, we have that if $\vp$ is valid on $F$ then $\vp$ is valid on
$F'$.
\end{theorem}

Two classes of frames $\cF_1,\cF_2$ are {\em validity-equivalent} with
respect a language $\cL$, denoted $\cF_1 \equiv_\cL \cF_2$, if for all
formulae $\vp \in \cL$, we have $\cF_1 \models \vp$ if and only if $\cF_2 \models
\vp$.

\begin{theorem} \label{thm:pmorph2} 
Suppose that $\cF_1$ and $ \cF_2$ are classes of frames 
such that for all $F\in \cF_1$ there exists a p-morphism from 
$F$ to a frame $F'\in \cF_2$, and conversely, 
for all $F\in \cF_2$ there exists a p-morphism from 
$F$ to a frame $F'\in \cF_1$. Then $\cF_1 \equiv_{\cL_n} \cF_2$
\end{theorem}

One further property of the language $\cL_n$ that will be of use to us
is the fact that satisfaction of a formula at a world depends only on
worlds connected to that world. Say that two worlds $w,w'$ of a frame
$F=(W, \sim_1, \dots, \sim_n)$ are {\em connected} if there exists a
finite sequence $w=w_0, \ldots, w_k=w'$ of worlds in $W$ such that for
$j=0\dots k-1$ we have $w_j\sim_i w_{j+1}$ for some $i$. Say that $F$
is {\em connected} if for all pairs of worlds $w,w'\in W$ are
connected. The {\em connected component} of $F$ containing a world $w$
is the frame $F_w = (W_w, \sim_1', \dots, \sim_n')$ where $W_w$ is
the set of worlds of $F$ connected to $w$, and each $\sim_i'$ is the
restriction of $\sim_i$ to $W_w$. Similarly, the connected component
of a model $M=(F,\pi)$ containing a world $w$ is the model $M_w =
(F_w,\pi')$ where $\pi'$ is the restriction of $\pi$ to $W_w$.  
The
model $M_w$ is also called the model generated by $w$ from $M$. 
The
following result (see, e.g., \cite{Hughes+Cresswell:1984} page 80)
makes precise the claim that satisfaction of a formula of $\cL_n$ at a
world depends only on connected worlds.

\begin{theorem} \label{thm:conn}
For all worlds $w$ of a model $M$ and for all formulae $\psi \in \cL_n$ we have
$M \models_w \phi$ if and only if $M_w \models_w \phi$.
\end{theorem}

A class of frames $\cF$ {\em corresponds} to a formula $\psi$ if 
for all frames $F$, we have $F \in \cF$ if and only if $F\models \psi$. 
We consider such correspondences between classes of frames and formulae
at several places in the paper, since they frequently indicate
that the formula can be used to obtain an axiomatization 
of the class of frames.

\subsection{Interpreted systems}

Interpreted systems are a model for distributed and multi-agent
systems proposed by Fagin, Halpern, Moses and Vardi
\cite{fhmv:rak,HalpernFagin85}, based on an earlier model of 
Halpern and Moses \cite{HalpernMoses90}. 
They provide a general theoretical framework within which it is
possible to model a variety of modes of communication, failure
properties of communication channels, and assumptions about
coordination such as synchrony and asynchrony. Its specific focus is
to enable states of knowledge to be ascribed to the agents in the
system, and to study the evolution of this knowledge as agents
communicate.  For discussion of axiomatic properties of this model see
\cite{fhmv:rak} and \cite{HalpernvanderMeydenVardi}. 
The key aspect of interpreted systems that allows knowledge to be
ascribed to agents is the notion of local state. Intuitively, the
local state of an agent captures the complete scope of the information
about the system that is accessible to the agent.  This may include
the values of its personal variables and data structures, its record of
prior communications, etc. The 
agents'
local states, together with 
a state of the environment within which they operate, determines
the global state of the system at any given time. 

Consider $n$ sets of local states, one for every agent of the 
system, and a set of states for the environment. We denote by $L_i$
the non-empty sets of local states possible for agent $i$, and by
$L_e$ the non-empty set of possible states for the
environment. Elements of $L_i$ will be denoted by $l_1, l_2,
\dots$. Elements of $L_e$ will be denoted by $l_e, \dots$.

\begin{definition} 
[System of global states] \label{is}
 \emph{A system of global states for $n$ agents} is a subset
  of a Cartesian product $L_e \times L_1\times \dots
  \times L_n$. 
An \emph{interpreted} system of global states is a 
pair $(S,\pi)$ where $S$ is a system of global states 
and $\pi: S\rightarrow 2^\atoms$ is an interpretation function for the
atoms.
\end{definition}

The reason for 
considering a subset is that some of the tuples
in
the Cartesian product might not be possible because of explicit
constraints present in the 
multi-agent system.  The framework of Fagin et al.\ \cite{fhmv:rak} models the
temporal evolution of a system by means of {\em runs}, which are
functions from the natural numbers to the set of global states. An
{\em interpreted system}, in their terminology, is a set of runs
over global states
together with a valuation for the atoms of the language on points of
these runs. We simplify this notion here, since we will deal initially
with an atemporal setting. However, we will consider a run-like
construct in Section~\ref{hbs-section}.

\newcommand\stof{F}

As shown in \cite{fhmv:rak}, 
interpreted systems 
can be used to ascribe knowledge to the agents by considering two
global states to be indistinguishable for an agent if its local state
is the same in the two global states. We formulate this here as a
mapping from systems of global states to Kripke frames.

\begin{definition}\label{stofr}
The function $\stof$ mapping systems of global states to Kripke frames
is defined as follows: if $S\subseteq L_e\times L_1\times \ldots
\times L_n$ is a set of global states for $n$ agents then $\stof(S)$
is the Kripke frame $(W, \sim_1,\dots, \sim_n)$, with $W=S$, and for
each $i= 1\ldots n$ the relation $\sim_i$ defined by $(l_1,\dots, l_n)
\sim_i (l'_1,\dots, l'_n)$ if $l_i=l'_i$.  The function $\stof$ 
is naturally extended to
map interpreted systems of global states to Kripke models as follows:
if $\stof(S) = (W, \sim_1,\dots, \sim_n)$ then $\stof(S,\pi) = (W,
\sim_1,\dots, \sim_n,\pi)$.
\end{definition}

Note that for all systems of global states $S$, the frame $\stof(S)$
is an equivalence frame.  Combined with the semantic interpretation of
the language $\cL_n$ on Kripke models, this mapping provides a way to
interpret $\cL_n$ on interpreted systems of global states (for $n$
agents).  We say that $\phi\in \cL_n$ is valid on an interpreted
system of global states $(S,\pi)$ if $\phi$ is valid on the model
$\stof(S,\pi)$.  Similarly, $\phi$ is valid on the system $S$ of
global states if $\phi$ is valid on $\stof(S,\pi)$ for all
interpretations $\pi$.

\subsection{Hypercube systems and full systems}

We now define two classes of systems of global states, {\em full
systems} and {\em hypercube systems}, that provide an intuitive model
for the initial situation in many systems of interest. These classes
both capture situations in which the agents do not have information
about each others' local states.  In hypercubes the environment is
assumed trivial, so there is no interesting correlation between the
agents' states and the environment.

\begin{definition} [Hypercube systems] \label{system}
  A \emph{hypercube system}, or \emph{hypercube}, is a Cartesian
  product $H=L_e \times L_1\times \dots \times L_n$, where
  $L_e$ is a singleton and 
  $L_1, \dots, L_n$ are
  non-empty sets. The class of hypercube systems is denoted by $\cal
  H$.
\end{definition}

In full systems, the agents may, however, have
some information about how their local state correlates with the state
of the environment. 

\begin{definition} [Full system] \label{fs}
  A system  $S \subseteq L_e \times L_1\times \dots \times L_n$ is {\em full} 
  if for every tuple $\la l_1, \ldots, l_n\ra \in L_1\times \dots \times L_n$
  there exists $s\in L_e$ such that $ \la s, l_1, \ldots, l_n\ra \in S$. 
  The class of full systems is denoted by $\cal FS$.
\end{definition}

Clearly, every hypercube is full. The converse is not true. The following 
example illustrates these definitions. 

\begin{example} \label{eg:cards}
Consider a card game with $n$ players and $n$ decks of cards.  At the
start of play, each player is dealt a hand of 12 cards, with player
$i$'s cards all drawn from deck $i$, where $1 \leq i \leq n$.  Each
player sees only their own hand, and not the hand of other players,
nor the undealt cards remaining in any of the decks.

The situation at the start of play may be described as a system of
global states as follows. Let $D$ be the set of all cards constituting
a deck.  A hand of 12 cards corresponds to a set $S \subseteq D$ with
$|S|= 12$.  Let $H$ be the set of such sets $S$. Then the set of
possible local states for each agent $i=1\ldots n$ is $L_i = H$. The
set of possible states of the environment is $L_e = \{ D \setminus
S~|~ S\in H\}^n$, i.e.  the set of arrangements in which each of the
$n$ decks has 12 cards removed. The set of global states of the system
is
\[S =  \{ \la s, h_1, \ldots, h_n \ra ~|~ h_i \in L_i~ 
\mbox{for $i=1\ldots n$ and}~ s = \la D\setminus h_1, \ldots
,D\setminus h_n\ra \}.
\]
This system is full, but not a hypercube. Using this modeling of the
initial state of the game, we may address questions concerning what a
player knows about the undealt cards. 
If the only issue of concern is a player's knowledge about the cards
that have been dealt, a more appropriate modeling of the game may be
to take $S = \{1\} \times H^n$ as the system of global states.  This
system is a hypercube.  (We note that the initial situation in most
cardgames, where players are dealt their hand from the same deck, is
neither a hypercube nor a full system. For example, a situation in
which two players both hold the ace of spades is not possible in such
a game.)
\end{example}

We will show in Section~\ref{hbs-section} that the applicability of the
class of full systems and hypercubes goes beyond that of modeling the
initial configurations of naturally occurring systems. 
We will define a dynamic framework that shows how agents' knowledge
changes over time, and illustrate it with extensions of the card-deck
example. 
Our first aim, however, will be to axiomatize the class of hypercubes
and full systems.  In order to use the tools of modal logics for this
aim we formally relate these classes of systems to several classes of
Kripke frames.

\section{Classes of frames corresponding to hypercubes and full systems}

In this section we identify a number of properties of frames that can
be used to characterize the frames corresponding to full systems and
hypercubes up to isomorphism. We also show that, somewhat
surprisingly, full systems and hypercubes generate precisely the same
set of valid formulae of the language $\cL_n$.  We obtain
this result by establishing the existence of p-morphisms between
frames in the classes of frames corresponding to these classes of
systems.

\subsection{Directed frames} 

We have seen above that every system of global states 
generates a frame. 
Our aim in this section is to characterize the frames generated by
hypercubes and by full systems.
The following result identifies some properties of the resulting
frames based on the properties of the system of global states.

\begin{lemma} \label{int}
Let $S$ be a system of global states, and 
let $\stof(S)=(W, \sim_1, \dots, \sim_n)$ be 
the frame defined from it by Definition \ref{stofr}. 
\begin{enumerate}
\item If $S$ is a hypercube then $F(S)$ is such that $\intersection_{i
\in A} {\sim_i}=id_W $;
\item If $S$ is full and $n\geq 2$ then $\stof(S)$ is connected. 
\item If $S$ is full then for any $w_1, \dots, w_n \in W$ there exists a $\overline{w}\in W$
such that $w_i \sim_i \overline{w}$ for each $i=1,\dots,n$.
\end{enumerate}
\end{lemma}
\begin{proof}
For (1), suppose that $S$ is a hypercube and consider any two elements
$w=(l_e, l_1,\dots, l_n)$, $w'=(l_e, l'_1, \dots, l'_n)$ in $W$ such
that $w (\intersection_{i \in A}
\sim_i) w' $.  (Note that the first component of these tuples must be the 
same if $S$ is a hypercube.)  Then for all $i$ in $A$,
$(l_e,l_1,\dots,l_n) \sim_i (l_e,l'_1,\dots, l'_n)$.  Therefore, by definition
of the relations $\sim_i$, for all $i$ in $A$ we have 
$l_i=l'_i$, that is $w=w'$.

For (2), suppose that $S$ is full and let $w=(l_e, l_1,\dots, l_n)$
and $w'=(l_e', l'_1, \dots, l'_n)$ be points in $S$.  Since $S$ is
full there exists $l_e''$ such that $w''=(l_e'', l_1, l_2'\dots,l'_n)$
is in $W$. Clearly $w \sim_1 w'' \sim_2 w'$. Thus, there is a 
path from $w$ to $w'$ of length two. 

For (3), suppose that $S$ is full and 
consider any $w_1=(l_e^1,l_1^1,\dots,l_n^1), \dots, 
w_n=(l_e^n,l_1^n,\dots,l_n^n)$. 
Since $S$ is full there exists $l_e$ such that 
$\overline{w}= (l_e,l_1^1,\dots,l_n^n)\in S$. By
Definition~\ref{stofr},
the world $\overline{w}$ is in $W$ and by construction for each
$i=1, \dots, n$, we have $w_i \sim_i \overline{w}$.
\end{proof}

This shows that Kripke frames that we build from the hypercubes and
full systems by means of the standard technique (\cite{fhmv:rak})
constitute a proper 
subclass of the class of equivalence frames.
We will show that the 
properties of Lemma~\ref{int} can be used to characterize the 
images of the hypercubes and full systems. 

We will say that a frame $(W, \sim_1, \dots, \sim_n)$ has the
{\em identity intersection property}, or is an {\em I frame}, if
$\intersection_{i \in A} {\sim_i}=id_W $. 
Similarly, we say that a frame is {\em directed}, or is a {\em
D frame}, if for any $w_1, \dots, w_n \in W$ there exists a
$\overline{w}\in W$ such that $w_i \sim_i \overline{w}$ for each
$i=1,\dots,n$.  We will also use combinations of these letters to
refer to frames satisfying several of these properties.  Thus,
directed equivalence frames with the identity intersection property
will be called {\em EDI frames}.  Similarly, we subscript $\cF$ by
these letters to indicate the class of frames have the corresponding
properties; thus $\cF_{EDI}$ denotes the class of EDI frames.
Lemma~\ref{int} states that the image of a full system under $\stof$
is an ED frame and since every hypercube is full, the image of a
hypercube is an EDI frame.

The converse of these properties is not true, e.g., it is not 
the case that every ED frame is the image of a hypercube. 
However, something very close to this is the case:

\newcommand{\cK}{{\cal K}}

\begin{lemma} \label{lem:dieqv} 
\begin{enumerate} 
\item For every ED frame $F$ there exists a full system $S$ 
such that $\stof(S) \equiv F$. 

\item For every EDI frame $F$ there exists a hypercube $S$ 
such that $\stof(S) \equiv F$.
\end{enumerate} 
\end{lemma} 

\begin{proof} 
We first show part (1). 
Let $F= ( W, \sim_1, \ldots, \sim_n )$ be an $ED$ frame.
Take $S \subseteq W \times W/{\sim_1} \times
\ldots \times W/{\sim_n}$ to be the set of tuples 
$\la w, [w]_{\sim_1} ,\ldots, [w]_{\sim_n} \ra $ 
where $w \in W$. We show that $S$ is a full system and  
such that $F(S)\equiv F$.
$\stof(S) \equiv F$. 

To see that $S$ is full, let $w_1, \ldots w_n \in W$. 
We show that there exists $w\in W$ such that 
$\la w, [w_1]_{\sim_1} ,\ldots, [w_n]_{\sim_n} \ra \in S$. 
Since $F$ is a $D$ frame, there exists a world $w$ such that $w \sim_i
w_i$ for each $i=1\ldots n$. Thus $\la w, [w_1]_{\sim_1} ,\ldots,
[w_n]_{\sim_n} \ra = \la w, [w]_{\sim_1} ,\ldots, [w]_{\sim_n} \ra \in
S$.  This shows that $S$ is full.

Write $\stof(S) = (S, \sim_1', \ldots , \sim_n')$. 
To show that $\stof(S) \equiv F$, 
define the mapping $h: S\rightarrow W$ by 
$h( \la w, [w]_{\sim_1} ,\ldots, [w]_{\sim_n} \ra) = w$. 
It is clear that $h$ is a bijection. Moreover, 
\[\begin{array}{rcl}
\la w_1, [w_1]_{\sim_1} ,\ldots, [w_1]_{\sim_n} \ra & \sim_i' & 
\la w_2, [w_2]_{\sim_1} ,\ldots, [w_2]_{\sim_n} \ra \\ 
& {\rm iff} & 
[w_1]_{\sim_i} = [w_2]_{\sim_i} \\ 
& {\rm iff} & w_1 \sim_i w_2 \\ 
& {\rm iff} & 
h(\la w_1, [w_1]_{\sim_1} ,\ldots, [w_1]_{\sim_n} \ra) \sim_i
h(\la w_2, [w_2]_{\sim_1} ,\ldots, [w_2]_{\sim_n} \ra).
\end{array}
\]
Thus, $h$ is a frame isomorphism, establishing $\stof(S) \equiv F$.
This completes the proof of part (1). 

For part (2), let $F= (W, \sim_1, \ldots, \sim_n)$ be an $EDI$ frame.
Define $S = \{ 1 \} \times W/{\sim_1} \times
\ldots \times W/{\sim_n}$.  Clearly $S$ is a hypercube. 
Write $\stof(S) = ( S, \sim_1', \ldots , \sim_n' )$. 
We show that $\stof(S) \equiv F$. 

Consider an element $\la 1, [w_1]_{\sim_1} ,\ldots, [w_n]_{\sim_n}
\ra$ of $S$.  Since $F$ is a D frame, there exists 
$w\in W$ such that $w\in [w_i]_{\sim_i}$ for each $i=1, \dots, n$.
Moreover, because $F$ is an I frame this $w$ is unique.  Define the
mapping $h: S\rightarrow W$ by taking $h( \la 1, [w_1]_{\sim_1}
,\ldots, [w_n]_{\sim_n} \ra)$ to be the unique $w$ such that $w\in
[w_i]_{\sim_i}$ for each $i=1, \dots, n$.  The mapping $h$ is surjective
because for each $w\in W$ we have $h( \la 1, [w]_{\sim_1} ,\ldots,
[w]_{\sim_n} \ra)=w$.  Moreover $h$ is injective because if $h( \la 1,
[w_1]_{\sim_1} ,\ldots, [w_n]_{\sim_n} \ra)= w= h( \la 1,
[w_1']_{\sim_1} ,\ldots, [w_n']_{\sim_n} \ra)$ then for each $i=1\dots
n$ we have that $w$ is in both $[w_i]_{\sim_i}$ and
$[w_i']_{\sim_i}$. Thus, these equivalence classes must be the
identical, and hence the tuples $\la 1, [w_1]_{\sim_1} ,\ldots, [w_n]_{\sim_n}
\ra$ and $ \la 1, [w_1']_{\sim_1} ,\ldots, [w_n']_{\sim_n} \ra$ 
are identical. 

It remains to show that $h$ has the homomorphism property. 
For this, note that by construction, 
for each $i=1, \dots, n$ we have 
$h(\la 1, [w_1]_{\sim_1} ,\ldots, [w_n]_{\sim_n} \ra) \sim_i w_i$. 
Thus if 
$ \la 1, [w_1]_{\sim_1} ,\ldots, [w_n]_{\sim_n} \ra \sim_i' 
\la 1, [w_1']_{\sim_1} ,\ldots, [w_n']_{\sim_n} \ra$ then 
$[w_i]_{\sim_i} = [w_i']_{\sim_i}$, 
hence 
$h(\la 1, [w_1]_{\sim_1} ,\ldots, [w_n]_{\sim_n} \ra) \sim_i w_i \sim_i w_i'
\sim_i
h(\la 1, [w_1']_{\sim_1} ,\ldots, [w_n']_{\sim_n} \ra)$.  Conversely,
suppose 
$u= h(\la 1, [w_1]_{\sim_1} ,\ldots, [w_n]_{\sim_n} \ra)$ 
and $v = h(\la 1,$ \linebreak $ [w_1']_{\sim_1} ,\ldots, [w_n']_{\sim_n} \ra)$ 
and $u\sim_i v$. 
By definition of $h$, $u\in [w_i]_{\sim_i}$ and 
$v\in [w_i']_{\sim_i}$. Since  $u\sim_i v$
it follows that $[w_i]_{\sim_i} = [w_i']_{\sim_i}$.
Thus, $ \la 1, [w_1]_{\sim_1} ,\ldots, [w_n]_{\sim_n} \ra \sim_i' 
\la 1, [w_1']_{\sim_1} ,\ldots, [w_n']_{\sim_n} \ra$. 
Thus, $h$ is a frame isomorphism, establishing $\stof(S) \equiv F$.
This completes the proof of part (2).
\end{proof}

Using Theorem~\ref{thm:fiso}, it follows from Lemma~\ref{int} and
Lemma~\ref{lem:dieqv} that from the point of view of the language
$\cL_n$, full systems and ED frames are equivalent, as are hypercubes
and EDI frames. Stated more precisely, we have the following.

\begin{theorem} \label{thm:framechar} 
$\stof({\cal H})
\equiv_{\cL_n} {\cal F}_{EDI}$ and $\stof({\cal FS}) \equiv_{\cL_n}
{\cal F}_{ED}$.
\end{theorem}

Our strategy 
for 
axiomatising these classes of systems will be to 
focus on the corresponding classes of frames instead. 
One approach to this would be to seek axioms that correspond 
to the properties D and I. This turns out not to be possible.

\begin{figure}
\epsfxsize = 2cm
\begin{center}
\input{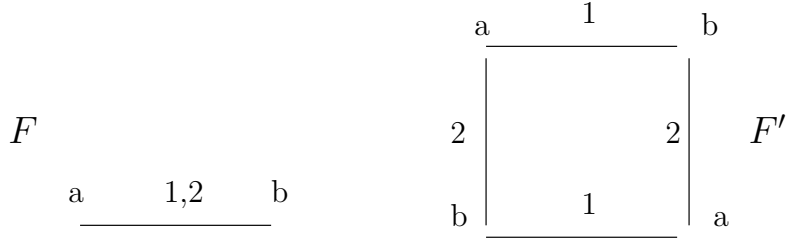}
\caption{Two p-morphic frames used in the proof of Lemma~\ref{corr-i}}\label{fig-i}
\end{center}
\end{figure}

\begin{lemma}\label{corr-i}
No modal formula corresponds to property I.
\end{lemma}
\begin{proof} Suppose the opposite and assume there is a formula $\phi$ that
  corresponds to property I. Consider the frame $F'$ in
  Figure~\ref{fig-i}. The frame $F'$ is an I frame, so $F'\sat \phi$.
  Consider now the frame $F$ and a function $p: F' \to F$ such that
  $p$ maps points in $F$ according to the names in the Figure. It is
  easy to see that $p$ is a p-morphism from $F'$ to $F$. Since
  p-morphisms preserve validity on frames (Theorem~\ref{thm:pmorph1})
  we have that $F \sat \phi$.
  But $F$ is not an I frame and we have a contradiction.
\end{proof}
Similar reasoning shows that the above holds even by restricting to
equivalence frames.

As an aside, we note that 
this result is very sensitive to the language under consideration. 
There are extensions of the language under which it fails. 
For example, consider a language containing an operator for distributed
knowledge \cite{JACM::FaginHV1992}. 
This operator is used to express the knowledge that the group 
of all agents would have if they pooled their information. 
Formally, if $\phi$ is a formula, then so is $D_A \phi$. 
The formula $D_A \phi$ is interpreted by
associating the relation ${\sim} = \intersection_{i \in A}\sim_i$ to the
operator $D_A$ in the standard Kripke-style interpretation, i.e., 
we define $M\models_w D_A \psi$ if 
$M\models_{w'}\psi$ for all $w' \sim w$. 
Using this operator, we can prove a correspondence result for
the intersection property. 

\begin{lemma} \label{da}
  An equivalence frame $F$ is an I frame if and only if $F\sat \phi \biimplies
  D_A \phi$.
\end{lemma}
\begin{proof}
Left to right.
Let $M$ be a model based on $F$ such that $M\sat_w \phi$. 
Since $\intersection_{i \in A} \sim_i= id_W$, then $M\sat_w D_A
\phi$. 
Analogously, suppose $M\sat_w D_A \phi$. Since 
$w(\intersection_{i \in A} \sim_i) w$
we have  $M\sat_w \phi$.\\
Right to left. Suppose $F\sat \phi \biimplies D_A \phi$ and  for all $i$
we have $w_1 \sim_i w_2$. Take a valuation $\pi$ such that
$p\in \pi(w)$ if and only if $w=w_1$. Since $F,\pi \sat_{w_1}p\biimplies D_Ap$ and
$(F,\pi) \sat_{w_1}p$, we have  $(F, \pi) \sat_{w_1}D_A p$ and so
$(F,\pi) \sat_{w_2}p$. But since $\pi(p)=\{w_1\}$, it must be that
$w_1=w_2$.
\end{proof}

This result suggests that the axiom $\phi \biimplies D_A \phi$ could
be used as part of an axiomatization of the class of EDI frames when
the language includes the distributed knowledge operator.  We are
concerned in this paper, however, with a weaker language.
Lemma~\ref{corr-i} suggests that it may be inappropriate to focus on
the identity intersection property in seeking to obtain the axiomatization.
Indeed, it turns out that this property has no impact on the 
set of valid formulae of $\cL_n$ in the context of interest
to us.  More precisely, we have the 
following result. 

\begin{theorem} \label{thm:edi-ed} 
$\cF_{EDI} \equiv_{\cL_n} \cF_{ED}$
\end{theorem}

To establish Theorem~\ref{thm:edi-ed}, we prove that any ED frame can
be seen as the target of a p-morphism from an EDI frame; the result
will then follow 
from 
Theorem~\ref{thm:pmorph2} using the fact that
p-morphisms between frames preserve validity and that the class of
DI frames is a subclass of the class of D frames. (Note that the
identity map on a frame is a frame isomorphism, hence a p-morphism.)
 
Consider an ED frame 
$F = (W, \sim_1, \dots, \sim_m)$. 
Write $\sim$ for the relation $\bigcap_{i=1\ldots m}
\sim_i$; since each of the $\sim_i$ is an equivalence relation, so is
$\sim$. The frame $F$ can then be viewed as the union of equivalence
classes of the relation $\sim$, which we call {\em clusters}.
Clusters containing more than a single point are sub-frames in which
property I clearly does not hold; in general a cluster may be infinite
in size.
 
\begin{figure}
\epsfxsize = 4cm
\begin{center}
\input{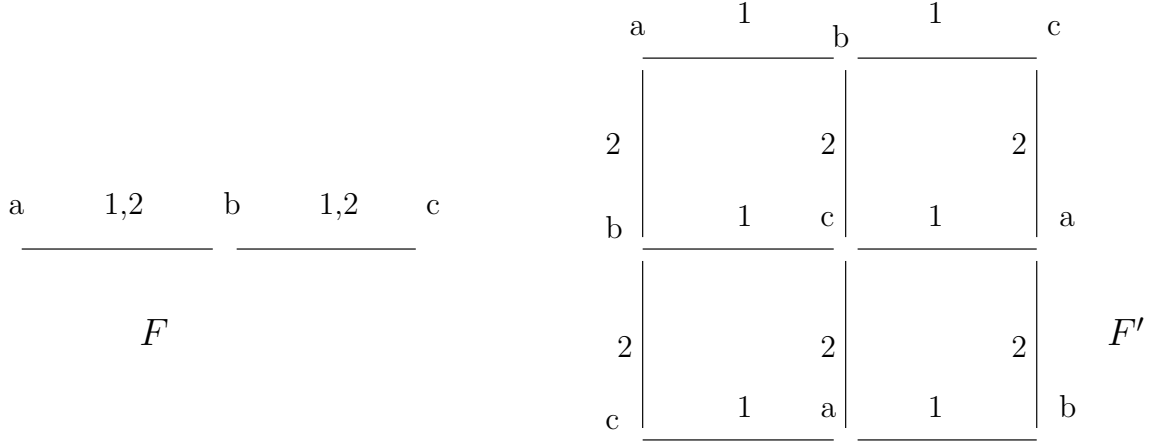}
\caption{A DI frame mapping a D frame via a p-morphism\label{di-figure}}
\end{center}
\end{figure}

If we want to construct an EDI frame that maps to a particular ED frame by a
p-morphism, one way is to replace every cluster of the ED frame with a
sub-frame that is EDI but that can still be mapped into the cluster.
Figure~\ref{di-figure} depicts the relatively simple case of an
equivalence frame $F$ composed by three points $a,b,c$ connected by
all the relations: $\sim_1, \sim_2$, in this case; $F$ clearly is ED
but not EI\footnote{
The relations are supposed to be the reflective transitive closure of
the the ones depicted in the figure.}.
The frame $F'$ on the right of the figure is an EDI frame; the names
of its points represent the targets of the p-morphism from $F'$ onto
$F$. So, for example the top left point of $F'$ is mapped onto $a$ of
$F$; the relations are mapped in the intuitive way. It is an easy
exercise to show that $F$ is indeed a p-morphic image of $F'$ and will
therefore validate every formula which is valid on $F'$.

The aim of the following is to define precisely how to build, given
any 
ED frame,
a new %
EDI frame in which every cluster is ``unpacked''
into an appropriate similar structure  and to define the relations
appropriately.

In order to achieve the above, we present two set theoretic results.
In Lemma~\ref{lemmax} we show that every infinite set $X$ can be seen
as the image of a product $X^m$ under a function $p$. Intuitively this
lemma will be used by taking the set $X$ as one of the clusters of an
EDI frame $F$, the function $p$ as the p-morphism and the product
$X^m$ (where $m$ is the number of relations on the frame) as the
sub-frame that will replace the cluster in the new frame $F'$.
Lemma~\ref{lemmac} extends the result of Lemma~\ref{lemmax} to
guarantee that even if the clusters differ in size it is always
possible to find a single sub-frame that can replace each of them.

We assume $m$ to be a natural number, such that $m \geq 2$.

\begin{lemma}\label{lemmax}
  Given any infinite set $X$, there exists a function
  $p: X^m \to X$ such that %
for all $i \in \{1,\dots, m\}$ and for all $u, x_i\in X$, there are
  $x_1,\dots,x_{i-1}, x_{i+1}, \dots, x_m \in X$, such that
  $p(x_1,\dots, x_m) = u$.
\end{lemma}
\begin{proof}
  Consider the set $T=\{\tau_{x,y} \mid x,y \in X\}$ of the
  transpositions of $X$, i.e.\ functions $\tau_{x,y}:X \to X;$ where
  $x,y\in X$, 
  such that if $z=x$ then $\tau_{x,y}(z)=y$, 
  if $z=y$ then $\tau_{x,y}(z)=x$,  and $\tau_{x,y}(z)=z$ otherwise. 
We have $| X | \leq |T| \leq | X
  \times X |$.  But by set theory (\cite{Lang} page 701 for example)
  $| X | = | X \times X |$, and so $| X | = |T|$. So, by induction, we
  have $| X^{m-1} | = | X | = | T |$. Call $f$ the bijection $f:
  X^{m-1} \to T$, and define $p( x_1, \dots, x_m )=f(x_1, \dots,
  x_{m-1})(x_m)$.  To prove the lemma
  holds we consider two cases: $i\neq m$ and $i=m$.

  For $i\neq m$, assume any $u\in X$, and any $x_i \in X$. Take any
  $x_j$ for $j\in \{1\ldots m-1\} \setminus \{i\}$. Then  $f(x_1, \dots,
  x_{m-1})$ is a transposition of $X$. So, there exists an $x_m \in X$
  such that $f( x_1, \dots, x_{m-1})(x_m)=u$. So $p(x_1, \dots, x_m)
  =u$. 

  For $i=m$, assume again any $u\in X$, and any $x_m \in X$.  Consider
  the transposition $\tau_{x_m,u}$; we have $\tau_{x_m,u}(x_m)=u$. But
  $\tau_{x_m,u}=f(x_1, \dots, x_{m-1})$ for some $x_1, \dots, x_{m-1}
  \in X$. So $p( x_1, \dots, x_m)=u$.
\end{proof}

Lemma~\ref{lemmax} induces a similar result for mappings from 
$X^m$ to sets whose cardinality
is smaller than $X$. 

\begin{lemma}\label{lemmac}
  Given any infinite set $X$, and a set $C\neq \emptyset$, such that
  $|C| \leq |X|$, there exists a function
  $p: X^m \to C$ such that the following holds for all $i \in \{1,\dots, m\}$: 
  for all $x_i\in X, u \in C$, there exist 
  $x_1,\dots,x_{i-1}, x_{i+1}, \dots, x_m \in X$, such that
  $p(x_1,\dots, x_m) = u$.
\end{lemma}
\begin{proof}
  Consider a set $T$ such that $C\cup T$ and $X$ have the same 
cardinality, and let $g$ be a bijection from $X$ to $(C \cup T)$. 
Then there is a function
  $p':(C \cup T)^m \to (C\cup T)$, satisfying the property expressed
  by Lemma~\ref{lemmax}. Define now a function $p'':(C\cup T) \to C$,
  such that $p''(x)=x \mbox{ if } x\in C$, otherwise $p''(x)= c$, where
  $c$ is any element in $C$. Define the function $p:X^m \to C$ 
  by $p(x_1, \dots, x_m) = p''(p'(g(x_1), \dots,g(x_m)))$. 
  We claim $p$ has the property required. 
  For, let $i\in \{1,\dots, m\}$ and take any $x_i \in X$
  and  $u\in C$. Then $g(x_i)\in (C \cup T)$,
  and so by Lemma~\ref{lemmax} there exist $c_1, \dots, c_{i-1}, 
  c_{i+1}, \dots, c_m \in C\cup T $,
  such that $p'(c_1, \dots,c_{i-1}, g(x_i), 
  c_{i+1}, \dots, c_m)=u$. Define $x_j = g^{-1}(c_j)$ 
  for $j\in \{1,\dots,m\} \setminus \{i\}$. We then have 
  $p(x_1, \dots,x_m) =  p''(p'(c_1, \dots,c_{i-1}, g(x_i), 
  c_{i+1}, \dots,$ \linebreak $ c_m)) = p''(u) = u$ since $u\in C$. 
\end{proof}

We rely on the two results above to define a function $p$ that maps
tuples $\langle c, x_1, \dots, x_m \rangle$ into $c$, where $c$ is a
cluster and $x_i \in X$, for some appropriate set $X$. The function $p$
is defined as in Lemma~\ref{lemmac} but it has an extra component for
the cluster.
 
\begin{corollary}\label{surj}
Let $\cC$ be a set of nonempty subsets of a set $W$. 
Then there exists a
  set $X$ and a function 
$p:\cC \times X^m \to W$ such that
\begin{enumerate} 
\item for all tuples $\langle c,x_1, \ldots, x_m\rangle$ we 
have $p(\langle c,x_1, \ldots, x_m\rangle) \in c$, and 
\item for all $c\in \cC$, for all $u\in c$, 
for all $i=1\ldots m$, and for all $x_i\in X$, 
for each $j\in \{1\ldots m\} \setminus \{i\}$ 
there exists $x_j\in X$, such that $p(\langle c, x_1\ldots x_m\rangle) =
u$. 
\end{enumerate}
\end{corollary}
\begin{proof}
Let $X$ be an infinite set with cardinality at least as great as 
the cardinality of any $c\in \cC$. 
This can be constructed by taking the union of these sets $c \in \cC$
or by considering the set of the natural numbers $X=\mathbb{N}$ if all
the sets $c \in \cC$ are finite. 
For each $c\in \cC$, let $p_c: X^m \to c$ be the function promised 
by Lemma~\ref{lemmac}. Define $p: \cC\times X^m \to W$ 
by $p(c, x_1,\dots,x_m) = p_c(x_1,\dots,x_m)$. 
It is immediate that this function has the required property. 
\end{proof}

\begin{theorem} \label{p-morphism}
Given any 
ED frame $F$, there exists an EDI frame $F'$,
and a p-morphism $p$, such that $p(F')=F$.
\end{theorem}
\begin{proof}  
  Let $F=(W,\sim_1, \ldots, \sim_m\rangle $ be a frame with $m$
  relations on its support set $W$. Write $\sim$ for the relation
  $\bigcap_{i=1\ldots m} \sim_i$. Since each of the $\sim_i$ is an
  equivalence relation, so is $\sim$.  Since the set of worlds $W$ of
  the frame $F$ is non-empty, it can be viewed as the union of the
  equivalence classes of the relation $\sim$, which we call {\em
    clusters}. Write  $\cC$ for the set of clusters of $F$. Consider the
  infinite set $X$ and a function $p$ as described in
  Corollary~\ref{surj}, and
define the frame $F'=(W', \sim_1',\ldots, \sim_m'\rangle$ as follows: 
\begin{itemize}
\item $W'= \cC \times X^m$, 
\item $\langle c, x_1, \dots, x_m
  \rangle \sim_i' \langle d,y_1, \dots, y_m \rangle$  if
  $x_i = y_i$ and 
  there exist worlds $u\in c$ and $v\in d$ such that 
  $u\sim_i v$.
\end{itemize}
We can prove that:

\begin{enumerate}
\item The frame $F'$ is EDI. 
\begin{proof}
  a) $F'$ is clearly an equivalence frame.

  b) We prove $F'$ satisfies property I.  Write $\sim'$ for
  $\bigcap_{i=1\ldots m}\sim_i'$.  Suppose $\langle c, x_1, \dots, x_m
  \rangle \sim' \langle d, y_1, \dots, y_m \rangle$.  Then for all
  $i=1\ldots m$ we have that $x_i=y_i$, and there exist $u_i\in c$ and
  $v_i\in d$ such that $u_i \sim_i v_i$. Since $c$ and $d$ are
  equivalence classes of $\sim$, it follows from the latter that
  $u_1 \sim v_1$, and consequently that 
  $c=d$. Thus, $\langle c, x_1, \dots, x_m
  \rangle = \langle d, y_1, \dots, y_m \rangle$. 

  c) We prove $F'$ satisfies property D. Consider $m$ tuples $\langle
  c_1, x^1_1, \dots, x^1_m \rangle,\dots,\langle c_m, x^m_1, \dots, x^m_m
  \rangle$ in $W'$. 
For each $i= i\ldots m$ let $u_i$ be a world in cluster $c_i$.
Since $F$ has property $D$, there exists a world $w$ such that 
$w\sim_i u_i$ for each $i=1\ldots m$. Let $c$ be the cluster 
containing $w$. Then, by construction, for each $i=1\ldots m$ 
we have $\langle c, x_1^1,\ldots, x_m^m\rangle 
\sim_i' \langle c_i, x_1^i,\ldots, x_m^i\rangle$.
\end{proof}
\item The function $p$ is a p-morphism from $F'$ to $F$. 
\begin{proof}
That the function $p$ 
is surjective follows from property (2) of Corollary~\ref{surj}. 

Next, we show that $p$ is a frame homomorphism.  Consider two tuples
$\langle c, x_1, \dots, x_m \rangle$, $\langle d, y_1, \dots, y_m
\rangle$ in $W'$ such that $\langle c, x_1, \dots, x_m
\rangle \sim'_i \langle d, y_1, \dots, y_m \rangle$. 
Then there exists $u\in c$ and $v\in d$ such that $u\sim_i v$.  By
property (1) of Corollary~\ref{surj}, we have $p(\langle c, x_1, \dots,
x_m \rangle) \sim_i u$ and $p(\langle d, y_1, \dots, y_m \rangle)
\sim_i v$. Since $\sim_i$ is an equivalence relation, it follows that
$p(\langle c, x_1, \dots, x_m \rangle)$ \linebreak $\sim_i p(\langle d, y_1, \dots,
y_m \rangle)$.

To show the backward simulation property, consider a tuple ${\bf
x}=\langle c, x_1, \dots, x_m \rangle$, and assume $p({\bf x}) \sim_i
w$ for some world $w$ of $F$. Let $d$ be the cluster containing $w$.
By Corollary~\ref{surj}(2), there exist $y_j$ for $j\neq i$ such that if
${\bf y}= \langle d, y_1, \ldots y_{i-1}, x_i, y_{i+1}, \ldots
y_m\rangle$, then $p({\bf y}) = w$. Since 
$p({\bf x}) \in c$ by Corollary~\ref{surj}(1), it is immediate that 
${\bf x} \sim_i' {\bf y}$. 
\end{proof}
\end{enumerate}
\end{proof}

This completes the proof of Theorem~\ref{thm:edi-ed}.  Since we
confine our attention in this paper to the language $\cL_n$, this
result, together with Theorem~\ref{thm:framechar}, shows that the set of
valid formulae for the class of full frames is the same as that for
the class of hypercubes.  Both sets of valid formulae are equal to the
set of formulae valid on ED frames. We now set about
attempting to axiomatize the latter. It turns out to be 
necessary to introduce one more class of frames in order to achieve this. 

\subsection{Weakly directed frames} 

In order to axiomatize the class of ED frames, we need to introduce
one more class of frames.  The reason for this is that the
directedness property does not naturally correspond to any formula of
$\cL_n$.

\begin{lemma}
No modal formula corresponds to n-directedness.
\end{lemma}
\begin{proof} Suppose the opposite and assume there is a formula $\phi$ that
  corresponds to n-directedness. Consider two disjoint frames, $F=(W,
  \sim_1,\dots,\sim_n)$ and $F'=(W',\sim_1',\dots,\sim_n')$, where $W
  \cap W' = \emptyset$, such that both $F$ and $F'$ are n-directed.
  the frame $F\cup F' = (W\cup W', \sim_1\cup
  \sim_1',\dots,\sim_n\cup\sim_n')$.  Since by assumption $F \sat
  \phi$ and $F' \sat \phi$, it follows that $F \cup F' \sat
  \phi$. (This is because satisfaction of a formula of $\cL_n$ at a
  world $w$ depends only on worlds connected to $w$
  (Theorem~\ref{thm:conn}).
  But, then $\phi$ is valid on a frame which
This is the opposite 
of 
what we assumed at the beginning.
\end{proof}

The problem here is rather superficial however. Any class of 
frames corresponding to a modal formula should be closed 
under disjoint unions. 
To address this problem, we define a slight weakening of the notion of
directedness. We will show that the class of frames satisfying this
weaker notion validates the same class of formulae.

\begin{definition}[Weak directedness] 
A frame $F=(W,\sim_1,\dots, \sim_n)$ is {\em weakly directed} 
when  for all worlds $w_0,w_1, \dots, w_n\in W$,  if 
for each $i=1, \dots, n$ there exists $j\in \{1, \dots, n\}$ such that 
 $w_0 \sim_j w_i$, then 
there exists a world $w$ such that $w_i \sim_i w$ for each $i=1, \dots, n$. 
\end{definition} 

That is, weak directness is like directedness in requiring the 
existence of a world $w$ such that $w_i \sim_i w$ for each $i=1, \dots, n$, 
but it does so only under the condition that the 
worlds $w_i$ are each connected to some world through a single step through 
one of the relations $\sim_j$. We use the notation ``WD'' to refer to the 
property of 
weak directedness. Thus, we write, e.g., 
$\cF_{EWD}$ for the class of weakly directed equivalence frames. 
Clearly, every directed frame is weakly
directed. Moreover, the class of weakly directed frames 
is easily seen to be closed under disjoint unions. Indeed,
this class of frames turns out to be the smallest class of 
frames containing the directed frames that is closed under 
disjoint unions. We first note the following.

\begin{lemma}\label{wdtod}
Every weakly-directed and connected equivalence frame 
is directed.
\end{lemma}
\begin{proof}
Suppose that $F = (W,\sim_1,\dots, \sim_n)$ is weakly directed and
connected. Let $w_1, \dots, w_n$ be any $n$ worlds in $W$.  We show
that there exists a world $w$ such that $w_i \sim_i w$ for each
$i=1, \dots, n$. Since $F$ is connected, the worlds $w_1, \ldots ,w_n$
are in the same connected component, and hence all connected to some
world $w'$. Since $F$  is an equivalence frame the relations $\sim_i$
are symmetric, so we may assume that for each $i$ there exists a path
directed from $w'$ to $w_i$.  We now claim that none of these paths
need to be any longer than one step, for if so, we can reduce their
length.  For, suppose without loss of generality that the path from
$w'$ to $w_1$ involves more than one step.  Write this path as $w'
\sim_{i_1} u \sim_{j_1} v \dots w_1$ and write the remaining paths as
$w' \sim_{i_2} w_2' \dots w_2$ to $w' \sim_{i_n} w_n' \dots w_n$.
Using weak directedness (and an ordering of the worlds $u, w_2'\dots
w_n'$ such that $u$ occurs in position $j_1$), we obtain a world $w''$
such that $u \sim_{j_1} w''$ and for each $k\neq 1$ we have $w_k'
\sim_{j_k} w''$ for some $j_k$. By symmetry of the
relations, we obtain paths from $w''$ to the worlds $w_i$. For $k\neq
1$ these paths are of the form $w'' \sim_{j_k} w_k'\dots w_k$ and have
the same length as the 
path connecting $w'$ to $w_k$.  For $k=1$ we have the path $w''
\sim_{j_1} u \sim_{j_1} v \dots w_1$, which can be shortened to $w''
\sim_{j_1} v \ldots w_1$ by transitivity of $\sim_{j_1}$.
This argument establishes that there exists a world $w'$ such that for
each $i=1, \dots, n$ we have $w' \sim_j w_i$ for some $j$.  Since $F$ is
weakly directed, it follows that there exists a world $w$ such that
$w_i \sim_i w$ for each $i=1, \dots, n$.
\end{proof}

We obtain two consequences of this result. First, the characterization 
of the weakly directed frames claimed above.

\begin{corollary} 
The class of weakly directed equivalence frames is the smallest class of 
equivalence directed frames that is closed under arbitrary disjoint
unions and isomorphism.
\end{corollary} 

\begin{proof} 
It is immediate from the definition that the class of weakly directed
equivalence frames contains the ED frames and is closed under disjoint
unions and isomorphism. To show that it is the smallest such class, we
show that any weakly directed equivalence frame is isomorphic to a
disjoint union of directed equivalence frames. For let $F$ be weakly
directed, and let 
$W'$ be a subset of the set of worlds of $F$
containing exactly one world from each connected component of $F$.
For each $w\in W$ let $F_w$ denote the connected component of $F$
containing $w$.  By Lemma~\ref{wdtod}, each $F_w$ is directed.
It is then possible to show that $F$ is isomorphic to
the disjoint union of the frames $F_w$ as $w$ ranges over 
$W'$.

\end{proof}

The second consequence of Lemma~\ref{wdtod} is the fact that 
the formulae of $\cL_n$ validated by the weakly equivalence directed frames
is the same as the set validated by the ED frames.

\begin{corollary} 
$\cF_{EWD} \equiv_{\cL_n} \cF_{ED}$
\end{corollary} 

\begin{proof} 
Since every ED frame is EWD, every formula valid on the EWD frames is
valid on the ED frames. Conversely, suppose that $\phi\in \cL_n$ is
not valid on some EWD frame $F$. Then there exists a valuation $\pi$
and a world $w$ such that $M \models_w \neg \phi$, where $M = (F,
\pi)$.  Let $M_w$ be the connected component of $M$ containing $w$ and 
$F_w$ the corresponding frame.  Then $M_w$ is a directed equivalence
model, and by Theorem~\ref{thm:conn} we have $M_w \models_w \neg
\phi$. Consequently, $\phi$ is not valid on the ED frame 
$F_w$. 
\end{proof} 

This result, together with the results of the preceding sections,
enables us to focus, in our quest for an axiomatization of the full
systems and hypercubes, on the class of weakly directed equivalence
frames.

\section{Axiomatization}

We are now ready to present an axiomatization of the full systems and
hypercubes with respect to $\cL_n$. The basis for the axiomatization
will be the property of weak directedness identified in the previous
section.

For convenience, we first introduce some notation and terminology.  We
will write $S \phi$ for the formula $\bigvee_{i=1, \dots, n} \Diamond_i
\phi$. Note that $M\models_w S \phi$ if there exists a world $w'$
such that $M\models_{w'} \phi$ and $w \sim_i w'$ for some $i$.
Intuitively, $S \phi$ asserts that at least one of the agents
$1, \dots, n$ considers $\phi$ possible.

For each $i=1, \dots, n$, we also define a formula to be $i$-local if it
is a boolean combination of formulae of the form $\b_i \phi$.
Intuitively, an $i$-local formula expresses a property of agent $i$'s
state of knowledge.  More precisely, we have the following fact, which
may be proved by a straightforward induction.

\begin{lemma} \label{lem:ilocal}
Let $\phi$ be an $i$-local formula in $\cL_n$, let $M$ be 
an equivalence model 
on $n$ agents, and let $w$ and $w'$ be two worlds of $M$ with 
$w\sim_i w'$. Then $M\models_w \phi$ if and only if $M\models_{w'} \phi$.
\end{lemma}

We analyze extensions of S5$_n$ with respect to the axiom schema:
$$
\left(\bigand_{i=1, \dots, n} S \phi_i\right) \implies SS\left(
\bigand_{i=1, \dots, n} \phi_i\right)
\eqno{{\bf WD}}
$$
where each $\phi_i$ is required to be an $i$-local formula. There is a
close relationship between this axiom, the property of weak
directedness and the property defining full systems.  Intuitively, the
axiom states that if there are $n$ worlds (each reachable in a single step
from the present world), such that the $i$-th world is one in which
agent $i$ is in a state of knowledge described by $\phi_i$, then there
exist a single world (reachable in two steps from the present) that realizes
these $n$ states of knowledge.  This intuitive
relationship may be made precise by the following correspondence
result:

\begin{lemma}\label{corr} For equivalence frames $F$, we have 
$F\sat {\bf WD}$ if and only if $F$ is weakly-directed.
\end{lemma}
\begin{proof}
We first show that if $F$ is a WD frame then $F\sat {\bf WD}$. For,
suppose that $\pi$ is an interpretation of $F$ and $w_0$ a world of
$F$ such that $(F,\pi) \models_{w_0} (\bigand_{i=1, \dots, n} S
\phi_i)$.  Then for each $i=1, \dots, n$ there exists a world $w_i$ such
that $w_0 \sim_{j_i} w_i$ for some $j_i$ and $(F,\pi)\models_{w_i}
\phi_i$. Since $F$ is weakly directed there exists a world $w$ such
that $w_i \sim_i w$ for each $i=1, \dots, n$.  By Lemma~\ref{lem:ilocal},
we have $(F,\pi)\models_{w} \bigand_{i=1, \dots, n} \phi_i$. Since $w_0
\sim_{j_1} w_1 \sim_1 w$, it follows that $(F,\pi)\models_{w_0} SS
\bigand_{i=1, \dots, n} \phi_i$.  This establishes $F\sat {\bf WD}$.

Conversely, suppose $F\sat {\bf WD}$.  We show that $F$ is weakly
directed.  Let $w_0,w_1, \dots, w_n$ be worlds of $F$ such that for
each $i=1, \dots, n$ there exists $j_i$ such that $w_0\sim_{j_i} w_i$. We
need to show that there exists a world $w$ such that $w_i \sim_i w$
for each $i=1, \dots, n$. To achieve this, let $p_1, \dots, p_n$ be $n$
distinct propositions and define the interpretation $\pi$ by $p_i \in
\pi(w)$ if and only if $w\sim_i w_i$, for each $i=1, \dots, n$. (The interpretation
$\pi$ may be defined arbitrarily on all other propositions.) Note that
we have $(F,\pi) \models_{w_0} \bigand_{i=1, \dots, n} S \b_i p_i$.
Since $F\sat {\bf WD}$, and each formula $\b_i p_i$ is $i$-local, it
follows that $(F,\pi) \models_{w_0} S S\bigand_{i=1, \dots, n}
\b_ip_i$. In particular, there exists a world $w$ such that $(F,\pi)
\models_{w} \bigand_{i=1, \dots, n} \b_ip_i$, hence 
$(F,\pi) \models_{w} \bigand_{i=1, \dots, n} p_i$. But, by definition of 
$\pi$, this means that $w_i\sim_i w$ for each $i=1, \dots, n$, as required. 
\end{proof}

The correspondence result above strongly indicates that the axiom {\bf
WD} can serve as a basis for an axiomatization of the weakly directed
equivalence frames. We now establish that this is indeed the case.
The proof will be by means of a standard technique for completeness
proofs in modal logic, namely the construction of a canonical model.
We now briefly review this technique
to fix the notation,
but refer the reader to \cite{bc:ml:i,Hughes+Cresswell:1984} for details.

A logic $L$ consists of a derivability relation $\proves_L$ typically
defined inductively using a basis of a set of axioms and closing under
a set of inference rules.  Given a logic $L$, a set of formulae
$\Gamma$ is {\em $L$-inconsistent} if there are formulae $\alpha_1,
\dots, \alpha_m \in \Gamma$, such that $\proves_L
\lnot (\alpha_1, \dots, \alpha_m)$, and {\em $L$-consistent} otherwise. A
set of formulae $\Gamma$ is {\em maximal} if for
every $\alpha$ of the language either $\alpha \in \Gamma$ or $\lnot
\alpha \in \Gamma$.  
Under appropriate conditions, 
it is possible to prove that every $L$-consistent set admits a maximal
$L$-consistent extension.

Given a multi-modal logic $L$, the canonical model $M_C^{L}=(W, R_1,
\dots, R_n, \pi) $ is a model for the logic $L$, built as follows. The
set $W$ is made of all the maximal $L$-consistent sets of formulae,
$R_i$ is a family of relations on $W^2$ defined by $wR_iw'$ if
$\forall{\alpha}$ $(\b_i \alpha
\in w$ implies $\alpha \in w')$. The interpretation $\pi$ for the atoms
is defined as $p \in \pi(w)$ if $p \in w$. 
For normal modal logics $L$ that are ``compact'' in the sense that 
all rules of inference have a 
finite number of antecedents, 
the canonical model has the property
that $M_C^L \sat \phi$ if and only if $\proves_L \phi$.

A logic $L$ is {\em sound} with respect to a class of frames $\fr$ if
$\proves_L \phi$ implies $\fr \sat \phi$. A logic $L$ is {\em
complete} with respect to a class of frames $\fr$ if $\fr \sat \phi$
implies $\proves_L \phi$.  Some logics are not only described by the
canonical model but also by the frame of the canonical model, called
the canonical frame. It can be proved that completeness of a logic $L$
with respect to a class of frames $\fr$ holds if the frame of the
canonical model is in $\fr$. Define the logic \S{} to be the logic
obtained from S5$_n$ by adding the axiom {\bf WD}. It is possible
to prove its completeness with respect to EWD frames. 

\begin{theorem}\label{comp-wd}
The logic \S{} is sound and complete for $\cL_n$ with respect to the
class of EWD frames.
\end{theorem}
\begin{proof}
  Soundness follows from what was proved in the first part of 
  Lemma \ref{corr} and the fact that all axioms and rules of S5$_n$ are
  sound for equivalence frames \cite{HalpernMoses90}.  To
  prove completeness we use the canonical model technique. It is easy
  to show that the frame $F_C^{\S}=(W, R_1, \dots, R_n)$ of the canonical
  model for \S\ is reflexive, symmetric, and transitive with
  respect to the $n$ relations. We prove it is also WD. 

Suppose that $w_0,w_1, \dots, w_n$ are worlds of $F_C^{\S}$
such that  $w_0 R_{j_i} w_i$, for $i=1,\dots, n$. 
Consider the set 
$$\Gamma=\union_{i=1}^n \{\phi : \b_{i} \phi \in w_i\}.$$ We show that
$\Gamma$ is \S-consistent.  It then follows by the maximal extension
theorem that there is a maximal \S-consistent extension $w$, which
satisfies $w_i R_i w$ by construction.  This will establish that the
frame is WD. To show $\Gamma$ is \S-consistent, we assume it is not
\S-consistent and obtain a contradiction.  It follows from the
assumption that for each $i = 1, \dots, n$ 
there are formulae $\alpha_1^i, \dots,
\alpha_{m_i}^i$, with $\b_{i} \alpha_1^j \in w_i$ for each 
$j=1\dots m_i$, such that 
$$
\proves_{\S} \lnot (\alpha_1^1\land \dots \land
\alpha_{m_1}^1 \land \dots \land \alpha_1^n \land \dots \land \alpha_{m_n}^n)
$$
Let us now call $\alpha_i = \land_{j=1}^{m_i} \alpha_j^i$. 
Note that by S5$_n$ reasoning, we have $\b_i \alpha_i \in w_i$. 
It follows that $\dia_{j_i} \b_i \alpha_i \in w_0$. 
(For else, $\b_{j_i} \neg \b_i \alpha_i\in w_0$, hence 
$\neg \b_i \alpha_i\in w_i$, contradicting consistency of 
$w_i$.) By propositional logic we obtain $S \b_i \alpha_i\in w_0$. 
Thus, $\bigand_{i=1, \dots, n} S \b_i \alpha_i\in w_0$. 
Now the formulae $\b_i \alpha_i$ are $i$-local, so using {\bf WD} 
it follows that 
$SS(\bigand_{i=1, \dots, n} \b_i \alpha_i) \in w_0$. 
By  S5$_n$ reasoning we get 
$SS(\bigand_{i=1, \dots, n} \alpha_i) \in w_0$.
But by S5$_n$ reasoning and the fact that 
$\proves_{\S} \lnot \bigand_{i=1, \dots, n} \alpha_i$ this leads to
 the conclusion that $w_0$ is inconsistent. This is the
contradiction promised. 
\end{proof}

Applying the equivalences with respect to 
$\cL_n$ established previously, we also obtain soundness and completeness 
with respect to several other semantics. 

\begin{corollary}\label{ax-hs}
The logic \S{} is sound and complete for $\cL_n$ with respect to 
\begin{enumerate} 
\item the class of full systems
\item the class of hypercube systems
\item the class of EDI frames 
\item the class of ED frames 
\end{enumerate} 
\end{corollary}

With Corollary~\ref{ax-hs} we have the axiomatization of full systems
and hypercubes systems that we aimed for. We remark that it can be
shown that several other axioms could be used for this result 
instead of {\bf WD}. For example, an analysis of the proofs 
of both the correspondence and completeness results reveals that 
the axiom 
$$
\left(\bigand_{i=1, \dots, n} S \b_i \phi_i\right) \implies SS\left(
\bigand_{i=1, \dots, n} \b_i\phi_i\right)
$$
(where the $\phi_i$ are not required to be $i$-local) also
suffices. This is not surprising, since for any $i$-local formula
$\phi$ it can be shown that $\phi \biimplies \b_i\phi$ is 
S5$_n$-valid. 
While the axiom {\bf WD} is compact when expressed using the operator
$S$, it is quite lengthy when expanded and involves considerable use
of disjunction. It is possible to show that {\bf WD} may be replaced
by certain other axioms which are less symmetrical, but which state
interactions between the agents' knowledge of a syntactically simpler
form than the expansion of {\bf WD}.
For a discussion of a number of 
alternative axioms 
that 
can be shown to be equivalent to {\bf WD}, we refer the reader to 
the thesis of Lomuscio~\cite{Lomuscio99}. 
One such alternative, for the case $n=2$,
has appeared in the literature before, as the axiom
 \[ \d_1\b_2 p \implies \b_2 \d_1 p\] 
due to Catach \cite{Catach88}, also discussed in \cite{popkorn}.

\begin{theorem}
  {\bf WD} in the case $n=2$ is $S5_2$-equivalent to Catach's axiom.
\end{theorem}
\begin{proof}
If $n=2$ then {\bf WD} is 
\[(\d_1\phi_1 \lor \d_2\phi_1) \land (\d_1\phi_2\lor \d_2\phi_2)
\implies
\bigvee_{i,j\in\{1,2\}}\d_i\d_j(\phi_1\land\phi_2)\]
(with $\phi_i$ $i$-local).

{\bf WD} to Catach: Put $\phi_1=\b_1\neg p$ and $\phi_2=\b_2 p$ (note
that these are 1-local and 2-local respectively). {\bf WD}
now becomes
\[(\d_1\b_1\neg p \lor \d_2\b_1\neg p) \land (\d_1\b_2 p \lor
\d_2\b_2 p) \implies \bot.\]
Now we drop the disjuncts  $\d_1\b_1\neg p$ and $\d_2\b_2 p$ (this
strengthens the antecedent and hence weakens the whole formula) to
obtain as a consequence
\[ \d_2\b_1\neg p \land \d_1\b_2 p \implies \bot,\]
which can be simply rearranged to obtain $\d_1\b_2 p \implies \b_2
\d_1 p$ as required.

Catach to {\bf WD}:
From $\d_1\b_2 p \implies \b_2 \d_1 p$ we want to obtain
\[(\d_1\phi_1 \lor \d_2\phi_1) \land (\d_1\phi_2\lor \d_2\phi_2)
\implies
\bigvee_{i,j\in\{1,2\}}\d_i\d_j(\phi_1\land\phi_2)\]
in the case that the $\phi_i$ are $i$-local.
Since the $\phi_i$ are $i$-local, we have
$\b_i\phi_i\biimplies\phi_i$. Assume
\[(\d_1\b_1\phi_1 \lor \d_2\b_1\phi_1) \land (\d_1\b_2\phi_2\lor \d_2\b_2\phi_2)\]
which, on distribution, is 
\[
(\d_1\b_1\phi_1 \land \d_1\b_2\phi_2)
\lor
(\d_1\b_1\phi_1 \land \d_2\b_2\phi_2)
\lor 
(\d_2\b_1\phi_1  \land \d_1\b_2\phi_2)
\lor 
(\d_2\b_1\phi_1  \land \d_2\b_2\phi_2)
\]
From each of these disjuncts, we will derive either
$\d_1\d_2(\phi_1\land\phi_2)$ or $\d_2\d_1(\phi_1\land\phi_2)$, thus
proving WD. The derivations are as follows:
\begin{enumerate}
\item From $(\d_1\b_1\phi_1 \land \d_1\b_2\phi_2)$, apply Catach's
  axiom together with uniform substitution to the second term to obtain
  $(\d_1\b_1\phi_1 \land \b_2\d_1\phi_2)$. Use the $S5_n$ axioms
  $\d_1\b_1\psi\biimplies\b_1\psi$ and $\b_2\d_1\psi\implies\d_1\psi$ to
  obtain $\b_1\phi_1\land\d_1\phi_2$. From this we deduce
  $\d_1(\phi_1\land\phi_2)$ and from the axiom T: $p \implies \d_2 p$
  and substitution we obtain $\d_2\d_1(\phi_1\land\phi_2)$.
\item From $(\d_1\b_1\phi_1 \land \d_2\b_2\phi_2)$: the first conjunct
  gives $\b_1\phi_1$, then $\phi_1$, then $\d_2\phi_1$ by $S5_n$
  axioms. The second conjunct gives $\b_2\phi_2$, so putting them
  together we have $\d_2\phi_1\land\b_2\phi_2$, from which we obtain
  $\d_2(\phi_1\land\phi_2)$ as a consequence, and hence
    $\d_1\d_2(\phi_1\land\phi_2)$.
\item From $(\d_2\b_1\phi_1 \land \d_1\b_2\phi_2)$, we obtain
  $\b_1\d_2\phi_1\land\d_1\b_2\phi_2$ by applying Catach to the first
  term. This now implies $\d_1(\d_2\phi_1\land\b_2\phi_2)$, which in
  turn implies $\d_1\d_2(\phi_1\land\phi_2)$.
\item From $(\d_2\b_1\phi_1 \land \d_2\b_2\phi_2)$: this case is
  similar to the first one.
\end{enumerate}

\end{proof}

\section{Decidability}

We now prove that the logic \S{} is decidable. In order to do that we
prove that the logic has the finite model property.

\begin{definition}
A logic $L$ is said to have the finite model property (or fmp in
short) if for any formula $\phi$, $\not \proves_L \phi$ implies that
there is a finite model $M$ for $L$ such that $M \not \sat \phi$.
\end{definition}

A logic can be proved to have the fmp in a number of different ways:
algebraically as in \cite{McKinsey}, \cite{Bergmann}, by the use of a
``mini-canonical'' model as in \cite{HughesCresswell96}, etc. Here we use
the another standard technique which is better suited for this case:
\emph{filtrations} (first presented in \cite{Lemmon}).

The idea of filtrations is the following. If a logic is complete, we know
that if a formula $\phi$ is a non-theorem of $L$ (i.e.\ if $\lnot
\phi$ is L-consistent), then $\phi$ is invalid on some model $M$ for
L. The model $M$ might be infinite. Filtrations enable us to produce a
model $M'$ from $M$, such that $M'$ is finite. If we can further prove
that $M'$ is also a model for L, then we have proved that the logic L
has the finite model property.

We formally proceed as follows.  Given a formula $\phi$, 
define the set $\Phi_\phi$ to be the set of formulae $\alpha$ 
that are either a sub-formula of $\phi$ or the negation of a 
 sub-formula of $\phi$. 
The set $\Phi_\phi$ is obviously finite for any formula $\phi$.
\begin{definition}\label{eq-points}
 Let $M$ be a model. 
 Two worlds $w,w'$ of $M$ 
 are equivalent with respect to $\Phi_\phi$
  (denoted $w \equiv_{\Phi_{\phi}} w'$,
   or simply $w \equiv w'$ if it is not ambiguous),  if for
 every  $\alpha \in \Phi_\phi$, we have $M\sat_w \alpha \mbox{ if and only if }
  M\sat_{w'} \alpha$.
\end{definition}

We can now define \emph{filtrations} as follows.

\begin{definition}
  Given a formula $\phi$ and a model $M=(W, R_1, \dots, R_n, \pi)$, a
  \emph{filtration} through $\Phi_\phi$ is a model $M'=( W', R'_1, \dots, R'_n,
  \pi')$ 
satisfying the following three properties:
\begin{itemize}
\item $W'= W/{\equiv_{\Phi_\phi}}$, where ${\equiv_{\Phi_\phi}}$ is
  the equivalence relation defined as in ~\ref{eq-points}.
\item For 
  each $i\in A$, the relation $R'_i$ is \emph{suitable}, i.e.\ it satisfies
  the two properties:
        \begin{enumerate}
        \item For all $[w_1],[w_2]\in W'$,  if there exists
              $u\in W$ such that
                $w_1 R_i u$ and $ u \equiv w_2$, then $[w_1] R'_i [w_2]$. 
              \item For all $[w_1],[w_2]\in W'$, if $[w_1] R'_i [w_2]$ then 
                 for all formulae ${\alpha}$ such that  $\b_i \alpha \in
                \Phi_{\phi}$, if  $M\sat_{w_1} \b_i
                \alpha$ then $M\sat_{w_2}\nolinebreak \alpha $.
        \end{enumerate}
\item  For any $p\in \atoms$, 
$p \in \pi'([w])$ if and only if $p \in \pi(w)$. 
\end{itemize}
\end{definition}
Note that 
a model $M'$ satisfying these conditions must be finite 
since $\Phi_\phi$ is finite, so the number of equivalence classes 
under $\equiv_{\Phi_\phi}$ is finite. 
Indeed, the number of worlds in $M'$ is at most $2^{|\phi|}$,

It can be proved by induction (see for example
\cite{Hughes+Cresswell:1984} page 139) that 
suitability of the relations $R_i'$ 
guarantees the validity of the following:

\begin{theorem} \label{truth-preservation-under-filtration}
  Given a model $M$, and any formula $\phi$, a filtration $M'$ of
  $M$ 
through $\Phi_\phi$ has the property that 
for any point $w\in W$ and and for any formula $\alpha \in \Phi$, 
we have $M'\sat_{[w]} \alpha$ 
if and only if $M\sat_w \alpha$
\end{theorem}

We now proceed to the case of interest here: the logic \S.
Consider the canonical model $M$
for \S. We
know (see Theorem \ref{comp-wd}) 
that $M$ is a weakly directed equivalence model. 
By Lemma~\ref{wdtod}, the model generated by any
point of $M$  is directed.  Consider any formula $\phi$. 
We consider the
model 
$M'$
defined as follows:

\begin{definition}\label{filtration-S5WD}
  Given a model $M$ and a formula $\phi$ define the model $M'=( W',
  \sim_1', \dots, \sim'_n, \pi')$
by 
\begin{itemize}
\item $W'= W/{\equiv_{\Phi_\phi}}$, where ${\equiv_{\Phi_\phi}}$ is
  the equivalence relation defined by Definition~\ref{eq-points}.
\item $[w_1]\sim'_i [w_2]$ if 
for all formulae $\alpha$ such that $\b_i \alpha \in
  \Phi_{\phi}$, we  have $M\sat_{w_1} \b_i \alpha$ if and only if 
  $M\sat_{w_2} \b_i \alpha$.
\item  
For any $p\in \atoms$, we have $p \in \pi'([w])$ if and only if  $p \in \pi(w)$.
\end{itemize}
\end{definition}

Indeed the model $M'$ defined by Definition~\ref{filtration-S5WD}
is a filtration as the following shows (stated in
\cite{Hughes+Cresswell:1984} page 145 for the mono-modal case).
\begin{lemma}
Given an
equivalence 
model $M$ and a formula $\phi$, the model $M'$ as described in
Definition~\ref{filtration-S5WD} is a filtration
of $M$ through $\Phi_\phi$. 
\end{lemma}
\begin{proof}%
All we need to prove is that the relations $\sim'_i$ are suitable.

 Property 1. Consider worlds $[w_1],[w_2] \in W'$ and world 
   $u\in W$ such that $w_1 \sim_i u$ 
  and $u\equiv w_2$. We need to prove that $[w_1] \sim'_i [w_2]$,
  i.e.\,  that for all formulae $\alpha$ such that 
  $\b_i \alpha \in \Phi_{\phi}$%
  we have 
  $M\sat_{w_1} \b_i \alpha$ if and only if $M\sat_{w_2} \b_i \alpha$.
  We prove it from left to right; the other direction is similar.
  Note that 
  $M\sat_{w_1} \b_i \alpha$ 
  if and only if $M\sat_{w_1} \b_i \b_i \alpha$
  because $M$ is an equivalence model; but $w_1 \sim_i u$ and so $M\sat_u
  \b_i \alpha$. But $\b_i \alpha \in \Phi$ and $w_2\equiv u$, so
  $M\sat_{w_2} \b_i \alpha$, which is what we wanted to prove. 

  Property 2. Consider worlds $[w_1],[w_2]\in W'$ such that 
  $[w_1] \sim'_i [w_2]$.
  This means that for all $\b_i \alpha \in \Phi$, we have $M\sat_{w_1} \b_i
  \alpha$ if and only if  $M\sat_{w_2} \b_i \alpha$. 
  Since $M$ is an equivalence model it follows that 
  $M\sat_{w_2} \alpha$.
\end{proof}

We now prove that the filtration defined above produces models for
\S. We first consider the effect of the filtration on directed 
models. 

\begin{lemma} \label{filtration-dir}
If $M$ is an equivalence directed model, 
then the model $M'$ defined in Definition~\ref{filtration-S5WD},
is also an equivalence directed model.
\end{lemma}
\begin{proof}
  We prove that $F'=( W', \sim_1, \dots, \sim'_n)$ is an ED frame. 
  The relations $\sim'_i$ are
  clearly equivalence relations. All it remains to show is that $F'$
  is directed. To do that, consider any $[w_1], \dots, [w_n] \in W'$. 
  Since $M$ is directed, there exists $w\in W$ such that 
  $w_i \sim_{i} 
w$ for $i=1,\dots, n$. But each $\sim'_i$ is
  suitable and so, by a consequence of property 1 of suitability we
  have that $[w_i] \sim'_{i} 
[w]$, for $i=1, \dots, n$.
  Therefore the frame $F'$ is directed. 
\end{proof}

We are finally in the position to prove fmp.

\begin{theorem}\label{thm:fmp} 
The logic $\S$ has the finite model property.
Indeed, every formula $\phi$ with a countermodel has a countermodel 
with at most $2^{|\phi|}$ worlds.
\end{theorem}
\begin{proof}
  Suppose $\not \proves \phi$. Since by the proof of
  Theorem~\ref{comp-wd} the logic $\S$ is canonical, the canonical model
  $M=( W, \sim_1, \dots,{\sim_n}, \pi)$ for $\S$ is an equivalence model, it is
  weakly-directed and there is a point $w \in W$, such that $M \sat_w
  \lnot \phi$. Consider the model $M_w$ generated by $w$. 
  By Theorem~\ref{thm:conn}, we have $M_w\sat_w \lnot \phi$. The
  model $M_w$ is clearly an equivalence model and, since it is
  connected it is also directed, by Lemma~\ref{wdtod}.  Consider now
  the filtration $M'$ of $M_w$ through $\Phi_\phi$ according to
  Definition~\ref{filtration-S5WD}; by Lemma~\ref{filtration-dir},
  $M'$ is an equivalence directed model and it is finite by
  construction because $\Phi_\phi$ is a finite set. But $M'$ is a
  filtration, and by Theorem~\ref{truth-preservation-under-filtration},
  $M'\sat_{[w]} \lnot \phi$, which is what we needed to prove.
  The bound on the size of $M'$ follows from the observation above. 
\end{proof}

\begin{corollary}
The logic $\S$ is decidable.
\end{corollary}
\begin{proof}
By Theorem~\ref{thm:fmp}, to check that $\phi$ is valid, it suffices 
suffices to check that $\phi$ has no countermodel with at most
$2^{|\phi|}$ worlds.
\end{proof}

Theorem~\ref{thm:fmp} is similar to a known result \cite{HalpernMoses92} for
the logic S5$_n$, for which an exponential size countermodel also
exists for every formula with a countermodel. (In the case of S5,
there exists a linear size model \cite{LadnerReif77}). We will leave open
the exact complexity of \S, but note that whereas the logic S5
is NP-complete \cite{LadnerReif77}, the logic S5$_n$ is known to be
PSPACE-complete \cite{HalpernMoses92}.  The upper bound for S5 is direct from
the existence of a linear size model, but it can be shown that, in a
precise sense, this technique does not work for S5$_n$. Instead, the
proof of the upper bound in the case of S5$_n$ is by means of a
tableau construction.  We will not attempt here to develop a similar
construction for \S.

\section{Homogeneous broadcast systems}\label{hbs-section}

\newcommand{\cg}{S}
\newcommand{\cgo}{I}

\newcommand\stt{{\rm \sigma}}

\newcommand\Or{\vee}

\newcommand\fin{{\it fin}}
\newcommand\init{{\it init}}
\newcommand\nll{\Lambda}

\newcommand\trees{{\cal T}}
\newcommand\obj{{\it obj}}

\newcommand\runs{{\it runs}}

\newcommand\red{{\it reduce}}

\newcommand\langc{{\cal L}_{n}^{C}}
\newcommand\langk{{\cal L}_{n}}

\newcommand{\vpo}{\varphi}
\newcommand{\vptw}{\varphi}

\newcommand{\ki}{K_i} 

\newcommand{\cR}{{\cal R}}
\newcommand{\cI}{{\cal I}}

\newcommand{\ba}{{\bf a}} 
\newcommand{\bP}{{\bf P}} 
\newcommand{\pg}{{\bf Pg}} 
\newcommand{\pgi}{{\bf Pg}_i}
\newcommand{\pgj}{{\bf Pg}_j}

\newcommand{\act}{{\it ACT}}
\newcommand{\goto}{{\rm goto}}
\newcommand{\assert}{{\rm assert}}

\newcommand\bA{{\bf A}}

\newcommand{\aseq}{{\bf a}}
\newcommand{\bj}{{\bf j}}
\newcommand{\bp}{{\bf p}}

\newcommand{\cko}{{\cal K}_1}
\newcommand{\ckn}{{\cal K}_n}
\newcommand{\cki}{{\cal K}_i}
\newcommand{\join}{\bowtie}
\newcommand{\joini}{\bowtie_i}

\newtheorem{assumption}{Assumption}[section]

Hypercubes were motivated above as an appropriate model for the
initial configuration of a 
multi-agent system, in which all agents are
ignorant of each other's local state. In this section we will show
that for a particular class of systems, homogeneous broadcast systems
with perfect recall, hypercubes are also an appropriate model of the
states of knowledge of agents that acquire information over time. In
this class of systems, all communication is by synchronous
broadcast, agents have perfect recall, and the agents'
knowledge in the initial configuration is characterized by a hypercube
system. We establish that in such systems, the agents' knowledge can
be characterized by a hypercube system not just at the initial time,
but also at all subsequent times.  It follows from this result that
the logic of knowledge in homogeneous broadcast systems can be
axiomatized by the logic $\S$ studied in the previous section.
Thus, the applicability of hypercubes as a model of agents' knowledge
extends beyond initial configurations.

This section is organized as follows. In Section~\ref{env} 
we 
describe \emph{environments}, a general model for the behavior of
agents and their interaction. This model may be used in a variety of
ways to ascribe a state of knowledge to the agents after a particular
sequence of events has occurred. We focus here on just one of the
possibilities, in which it is assumed that agents have {\em perfect
recall} of their observations.  In Section~\ref{be}, we define
broadcast environments, a special case of this general model that
constrains all communication between agents to be by synchronous
broadcast. Section~\ref{hbs} considers the special case of homogeneous
broadcast environments, establishes the connection between the systems
generated by these environments and hypercubes.

\subsection{Environments}\label{env}

In the model of Halpern and Moses \cite{HalpernMoses90}, a distributed system
corresponds to a set of runs, where each run constitutes a history
that identifies at each point of time a state of the environment and a
local state for each agent. This model is perhaps overly general,
since in practice one is interested in the particular sets of runs
that are generated by executing a given program, or {\em protocol},
within a given communication architecture.  A formal framework to
capture this idea, {\em contexts}, was defined by Fagin et al.\
\cite{FHMV:97,fhmv:rak}.  In this section we briefly recall a variant of
this framework, {\em environments}, from \cite{Mey-tark96}. (We refer
the reader to \cite{Mey-tark96,fhmv:rak} for more extensive motivation and
examples.) Compared to contexts, environments admit an additional
degree of freedom by allowing knowledge to be interpreted in different
ways in the same set of runs. We focus here on a particular interpretation, 
based on the assumption that agents have perfect recall. 
 We describe how executing a protocol in
an environment with respect to an interpretation of knowledge
determines a Kripke structure that ascribes a state of knowledge to
the agents after the occurrence of a particular sequence of events. In
Section~\ref{be}, we will present a special case of this model that
defines a particular architecture in which agents communicate by
synchronous broadcast.

\newcommand{\Prop}{\atoms}

For the definition of environment, we assume a set $A=\{0,
1,\dots,n\}$ of agents.  We also assume that for each agent $i\in A$,
there is a non-empty set $\act_i$, representing the set of
\emph{actions} that may be performed by agent $i$.  A \emph{joint
action} is defined to be a tuple $ \la a_0, \dots, a_n \ra \in \act_0
\times \dots \times \act_n$. We write $\act$ for the set of joint
actions. As before, we assume a set $\Prop$ of propositional variables
of the language.

In the following definitions, agent $0$ will play a role somewhat
different from the other agents. Intuitively, it is intended that 
agent 0 be used to model aspects of the context, or communication
architecture, within which the other agents operate.  The actions of agent
0 correspond to nondeterministic behavior of this context.
In applications of the framework, the architecture is typically
fixed, and one is interested in designing programs for the behavior 
of agents $1\ldots n$.

\begin{definition}[Environment]\label{Environment}
An \emph{interpreted environment} is a
tuple of the form $E= \la \cg, \cgo, P_0,$ \linebreak $\tau , O ,V\ra$ where the
components are defined as follows:
\begin{itemize}

\item $\cg$ is a  set of {\em
states of the environment}. Intuitively, states of the environment may
encode such information as messages in transit, failure of components,
etc.

\item  $\cgo$ is a subset of $\cg$, representing the possible 
{\em initial states\/} of the environment. 

\item $P_0:\cg \rightarrow {\cal P}(ACT_0)$ is a function,
called the {\em protocol of the environment,} mapping 
states to subsets of the set $\act_0$ of actions performable by the
environment.  Intuitively, $P_0(s)$ represents the set of actions that
may be performed by the environment when the system is in state $s$.

\item $\tau$ is a function mapping joint actions $\bj \in \act$ to 
state transition functions $\tau(\bj):\cg \rightarrow \cg$. 
Intuitively, when the joint action $\bj$ is performed
in the state $s$, the resulting state of  the environment is 
$\tau(\bj)(s)$. 

\item $O$ is a function from $\cg$ to ${\cal O}^n$  for 
some set ${\cal O}$. For each $i=1,\dots,n$, the function $O_i$ mapping
$s\in \cg$ to the $i$th component of $O(s)$, is called the {\em
observation function of agent $i$.}  Intuitively, $O_i(s)$ represents
the {\em observation} of agent $i$ in the state $s$.

\item $V:\cg\times \Prop\rightarrow \{0,1\}$ is a valuation, 
assigning a truth value $V(s,p)$ in each state $s$ 
to each atomic proposition $p\in \Prop$. 
\end{itemize}
\end{definition}

A {\em trace} of an environment $E$ is a {\em finite} sequence $s_0
\ldots s_m$ of states such that $s_0\in I$ and for all $k=0\ldots m-1$ there exists a
joint action $\bj = \la a_0, a_1,\ldots, a_n\ra$ such that $s_{k+1}
= \tau(\bj)(s_k)$ and $a_0 \in P_0(s_k)$.  We write $\fin(r)$
for the final state of a trace $r$.  

Intuitively, the traces of an environment correspond to the finite
histories that may be obtained from some behavior of the agents in
that environment.\footnote{One could also define \emph{runs} of the
environment, which are infinite sequences of states satisfying the
same constraint on state transitions. This would correspond more
closely to the framework of \cite{fhmv:rak}. Runs are essential when one
is interested in languages containing temporal operators, but there is
a precise sense in which it suffices to work with traces when only modal
operators for knowledge are of interest, as in the present paper. See
the appendix of \cite{Meyden1998} for a discussion of this issue.}  Note
that the nondeterministic choices of action made by the environment
itself are constrained by the protocol of the environment, and that
these choices are determined at each step from the state of the
environment.  On the other hand, the notion of trace assumes that the
choices of action of agents $1\ldots n$ are unconstrained. In
practice, we wish these agents to behave according to some program
(perhaps nondeterministic), that determines their choice of next
possible action as some function of the observations that they have
made.  The following definition captures this intuition.

\begin{definition}[Perfect Recall]
The {\em perfect recall local state} of agent $i=1\ldots n$ in a trace
$r = s_0\ldots s_m$, denoted $\{r\}_i$, is defined to be the sequence
$O_i(s_0)\ldots O_i(s_m)$ of observations made by the agent in the
trace.

A {\em perfect recall protocol} for agent $i=1\ldots n$ is a function
$P_i$ mapping each sequence of observations in ${\cal O}^*$ to a non-empty
subset of $\act_i$. A {\em joint perfect recall protocol}
is a tuple $\bP = \la P_1, \ldots , P_n\ra$ consisting of a perfect
recall protocol $P_i$ for each agent $i=1\ldots n$. We write $\bP_i$
for $P_i$ when $\bP$ is given.
\end{definition}

Protocols  specify the possible choices of next
action of the agents, given a certain history of events, as follows.
For each agent $i=1\ldots n$, we say that an action $a_i \in \act_i$
is {\em enabled} with respect to a protocol $\bP$ at a trace $r$ of
$E$ if $a_i \in \bP_i(\{r\}_i)$.  An action $a_0$ of the environment
is enabled at $r$ if $a_0 \in P_0(\fin(r))$. A joint action is enabled
at $r$ with respect to a protocol $\bP$ if each of its components is
enabled at $r$.

We obtain the traces that result when agents execute a joint protocol
in an environment as follows.  Define a trace $s_0\ldots s_m$ of $E$
to be {\em consistent} with a joint protocol $\bP$ if for each $k<m$,
there exists a joint action $\bj$ enabled at $s_0\ldots s_k$ with
respect to $\bP$, such that $\tau(\ba)(s_k) = s_{k+1}$.  We are now in
a position to describe the 
frame  that captures the agents'
states of knowledge when they execute a protocol in an
environment.

\begin{definition}[Perfect recall frame derived from a protocol and environment]\label{prframe}
Given \linebreak an environment $E$ and a joint protocol $\bP$, the \emph{perfect
recall frame} derived from $E$ and $\bP$ is the structure
$F_{E,\bP} =
(W, \sim_1, \ldots, \sim_n)$, where:
\begin{itemize} 
\item $W$ is  the set of all traces of $E$
consistent with $\bP$, %
\item $\sim_i$ is the 
binary relation on $W$ defined by 
 $r\sim_i r'$ if $\{r\}_i = \{r'\}_i$, for  each agent $i=1 \ldots n$.
\end{itemize} 
\end{definition}

Intuitively, because $W$ contains only traces of $E$ consistent with $\bP$,  
this 
frame encodes the assumption that it is common knowledge
amongst the agents that the environment in which they are operating is
$E$ and that the protocol they are running is $\bP$. Moreover, the
accessibility relations $\sim_i$ expressing agents' knowledge are
defined in a way that corresponds to assuming that agents have
perfect recall of their observations.  The relations $\sim_i$ could
have been defined in many different ways: for example, it is
meaningful to consider instead the relations $\approx_i$ defined by $r
\approx_i r'$ if 
$O_i(\fin(r)) = O_i(\fin(r'))$.  This would correspond to the
assumption that agents are only aware of their most recent
observation. The assumption of perfect recall we work with in this
paper is frequently made in the literature because it amounts to
assuming that agents make optimal use of the information to which they
are exposed. This assumption is essential for the derivation of lower
bounds and impossibility results and the synthesis of optimal
protocols \cite{HalpernMoses90,MosesTuttle88,Halpern+90}.

\subsection{Broadcast environments}\label{be}

In this subsection we define \emph{broadcast environments} (BE), a
special case of the formalism described in
Section~\ref{env}. Broadcast environments model situations in which
all communication is by synchronous broadcast.  Examples of this are
systems in which agents communicate by means of a shared bus,
by writing tokens onto a shared blackboard
\cite{Nii86}
and in face to face conversation.
Other examples are classical puzzles such as the wise men, or muddy
children puzzle \cite{Moses+86}, and a variety of games of incomplete
information, including battleships, Stratego and Bridge.  Broadcast
environments have been considered previously in \cite{Meyden96b}.

To define broadcast environments, we need to impose a number of constraints
on the components making up the definition of environments given in
the previous section. We do so here in a way that slightly simplifies
the model in \cite{Meyden96b}, eliminating some features that will
be irrelevant in the context of homogeneous broadcast environments.  The
intuition we wish to capture is that each agent holds some private
information, which is unobservable to all other agents. The actions
taken by the agents will have two types of effects: they will update this
private information, and simultaneously broadcast some information to all
the other agents.

The actions performed by agents in broadcast environments  have two
components: an {\em internal\/} component and an {\em external\/}
component.  The internal component of an agent's action will affect
only the agent's 
private state, and will be unobservable
to the other agents. On the other hand, the external component will be
observable to all agents, but it will affect only the state of the
environment. 
\begin{assumption}[BE Actions] \label{ass:act} 
For each $i=0\ldots n$
there exists a set $A_i$ of {\em external actions} and a set $B_i$ of
{\em internal actions}. All the sets $A_i$ contain the special
``null'' action $\epsilon$. For $i=0\ldots n$, the set
$\act_i$ of actions of agent $i$  consists of the pairs 
$a\cdot b$ where $a\in A_i$ and $b\in B_i$.
\end{assumption} 
The role of the null action is to allow for a uniform representation
of initial states (see Assumption~\ref{ass:init} below).  It follows
from 
Assumption~\ref{ass:act} 
and the definitions of the previous section
that the set of the joint actions $\act$ in a broadcast environment
consists of the tuples of the form ${\bf j} =
\la a_0\cdot b_0, a_1\cdot b_1,\ldots ,a_n\cdot b_n\ra $ where
$a_i\cdot b_i \in \act_i$ for each $i= 0\ldots n$.  We define
$\ba({\bf j})$ to be the component $
\la a_0,a_1,a_2,\ldots, a_n \ra$, and call this the \emph{joint external 
component} of ${\bf j}$. We write $\bA$ for the set $A_0 \times \ldots
\times A_n$ of joint external actions.

To represent the private information held by agents, we assume that
for each agent $i=0\ldots n$ there exists a set $S_i$ of {\em
instantaneous private states}. Intuitively, for $i=1\ldots n$ the states $S_i$
represent the information observable by agent $i$
only. In the case $i=0$, the states $S_0$ represent that part
of the environment's state which is observable to no agent.
\begin{assumption}[BE States]\label{ass:states}
The set  of  
states $\cg$ of a broadcast environment is required to consist of tuples of the
form $\la a_0,\ldots, a_n;p_0,\ldots ,p_n\ra$, where for each
$i=0\ldots n$, the component $a_i\in A_i$ is an external action of agent $i$
and the component $p_i\in S_i$ is a private state of agent
$i$.
\end{assumption} 
Intuitively, a tuple $\la a_0,\ldots, a_n;p_0,\ldots ,p_n\ra$ models a
situation in which each agent $i$ is in the instantaneous private state
$p_i$, and in which $a_i$ is the most recent external action
performed by agent $i$.
We define the {\em joint private state at} $s$ to be the tuple ${\bf
p}(s) = \la p_0,\ldots, p_n\ra$, and {\em agent $i$'s private state
at} $s$, denoted ${\bf p}_i(s)$, to be the private state $p_i$.
If $s = \la a_0,\ldots, a_n;p_0,\ldots ,p_n\ra$ is a
state then we define the {\em joint external action
at} $s$, denoted $\ba(s)$, to be the tuple $\la a_0,\ldots, a_n\ra$.

Clearly, in initial states it does not make sense to talk of a most
recent external action.  This motivates the following. 
\begin{assumption}[BE initial states] \label{ass:init}
The set of initial
states $\cgo$ of a broadcast environment contains only states $\la
a_0,\ldots, a_n;p_0,\ldots ,p_n\ra$ with $a_i = \epsilon$ for all
$i=0\ldots n$. 
\end{assumption} 
The definition of broadcast environment allows the set
$\cgo$ of initial states to be any nonempty set of states of this
form. As we will see later, homogeneous broadcast environments
restrict the possible sets of initial states.

In a broadcast environment, agents are aware of their own private state,
and also of the external actions performed by all agents. 
All communication between agents will be by means of the
external actions.  This constraint is model-led by the definition of the
agents' observations.  
\begin{assumption}[BE observations]\label{ass:obs} 
For $i=1\ldots n$, we require agent $i$'s 
observation function $O_i$ to be given by 
$O_i(\la a_0,\ldots,a_n;p_0,\ldots, 
p_n\ra) = \la a_0,\ldots,a_n;p_i \ra$. 
\end{assumption} 
 That is, in a given state, an agent's observation consists of the
external component of the joint action producing that state, and the
agent's private state.\footnote{ This is a slight simplification of
the definition in \cite{Meyden96b}, eliminating an extra component
that is incompatible with homogeneity.}

It will be convenient in what follows to introduce an observation
function for agent 0, similarly defined by $O_0(\la
a_0,\ldots,a_n;p_0,\ldots, p_n\ra) = \la
a_0,\ldots,a_n;p_0\ra$. Moreover, we obtain using this observation
function an equivalence relation $\sim_0$ on traces, defined exactly 
as the relations $\sim_i$.

One of the effects of performing a joint action in a broadcast
environment is that each agent updates its private state in a way that
depends on its internal action and the joint external action
simultaneously being performed. In addition to this, the joint
external action will be recorded in the resulting state.
\begin{assumption}[BE transitions]\label{ass:trans} 
For each agent $i=0\ldots n$ there exists a
private action interpretation function $\tau_i:\bA \times B_i
\rightarrow (S_i \rightarrow S_i)$. 
The joint action interpretation function $\tau:\act\rightarrow
(\cg\rightarrow\cg)$ of a broadcast environment is obtained from the
private action interpretation functions as follows. For each joint
action $\bj =
\la a_0\cdot b_0, \ldots, a_n\cdot b_n\ra$,  
the transition function $\tau({\bf
j})$ maps a state $s= \la a_0',\ldots,
a_n';p_0,\ldots ,p_n\ra$ to
\[\tau({\bf j})(s) = 
\la a_0, \ldots, a_n;\tau_0(\ba(\bj),b_0)(p_0), \ldots ,\tau_n(\ba(\bj),b_n)(p_n)\ra.\]
\end{assumption} 
That is, for each joint external action $\ba\in \bA$ and internal
action $b_i\in B_i$, the function $\tau_i(\ba, b_i):S_i\rightarrow
S_i$ is a private state transition function, intuitively representing
the effect on  agent $i$'s private states of performing the internal
action $b_i$ when the joint external action $\ba$ is being
simultaneously performed.  The state of the environment resulting from
a joint action records the external component of the joint action, and
updates each agent's private state using its private
action interpretation function.

Finally, we require that the protocol $P_0$ of the environment depend only
upon its private state and the most recent external action.
\begin{assumption}[BE protocol]\label{ass:prot}
If $s$ and $t$ are states with $O_0(s)= O_0(t)$ then  $P_0(s) = P_0(t)$. 
\end{assumption} 
The propositional constants $\Prop$ of a broadcast environment are
allowed to describe any property of the global states $\cg$, so we do
not make any assumption on the valuation $V$.

We may now state the main definition of this section. 
\begin{definition}[Broadcast environment] 
A {\em broadcast environment\/} is 
an environment satisfying 
assumptions~\ref{ass:act}---\ref{ass:prot}. 
\end{definition} 
In broadcast environments, agents' mutual knowledge can be shown to
have a particularly simple structure \cite{Meyden96b}. Intuitively,
this is because broadcast communication maintains a high degree of
common knowledge. The following section provides an illustration of
this point in a special case of broadcast environments.

\begin{example}\label{eg:be}
We illustrate the definitions so far by means of a simple card game.
Let us imagine an initial situation set up as described in
Example~\ref{eg:cards}, and take $n=2$.  Thus  each of the two players 
has 12 cards from the respective deck.  The game now proceeds as follows:
at each move, provided the hands are not empty, both players select a
card from their hand, and place it face up on the table, where it
remains until the next move, when it is returned to its deck, so that
it is no longer visible.  (Of course, a player with perfect recall
will remember what cards have been played.) The objective of the game
does not concern us here (it may be to play a card with a value
greater than that picked by the opponent, for example). Note that the
players play in parallel (synchronously).  We model this game as a
broadcast system together with a joint protocol.

The private states of agent $i\geq 1$ are now of the form $h\subset
D$, where $D$ is a deck and $h$ is a set of 12 or fewer cards.  As in
Example~\ref{eg:cards}, we will consider two modelings for the
private states of agent $0$. The first modeling, which we call the
{\em simple} modeling, takes the states of the environment to be the
singleton set $\{1\}$. This is appropriate when we wish to analyze the
players' knowledge about each other's hands.  If we also wish to
consider the players' knowledge about the cards in the respective
decks, then we use the {\em rich} modeling, in which we take agent
$0$'s private states to be tuples $\la D_1, D_2, f_1,f_2\ra$,
where $D_i\subseteq D$ represent the cards in the $i$-th deck, as
before, and $f_i \subset D$ contains at most one card, the card most
recently placed face up by player $i$. (In the initial states, we will
have $f_i$ empty.)

In this example, the environment is passive, so we may take the
actions of agent $0$ to consist of the single pair $\epsilon\cdot
\epsilon$ only, and for each state $s$ the environment's protocol
returns $P_0(s) = \{\epsilon\cdot \epsilon\}$.  All the actions of the
remaining agents are observable, so we may take the set of internal
actions $B_i$ to be $\{\epsilon\}$ in each case, and the set of
external actions $A_i$ to be equal to the set of subsets of $D$ with
at most one element.  Intuitively, $c \in A_i$ corresponds to the
action of playing the cards in $c$, so an empty set represents the
action of playing no card. Thus, the set of joint actions is the set
of tuples of the form
\[ \bj =  \la \epsilon\cdot \epsilon,~ c_1\cdot \epsilon,~ c_2\cdot \epsilon \ra\] 
where $c_1,c_2 \in A_i$, and the corresponding joint external action has the 
form $\ba(\bj) = \la \epsilon, c_1,c_2 \ra$.

Given the above, we also see that
a state of the system is a tuple of the form 
\[ s= \Big \la \epsilon\cdot \epsilon,~ c_1\cdot \epsilon,~ c_2\cdot \epsilon; ~ 
   p_0, h_1,h_2 \Big \ra\] where, for $i=1,2$, we have that $c_i
\subset D$ is a set of cards with at most one element, and $h_i$ is a
set of twelve or fewer cards.  In the simple modeling we have $p_0=1$;
in the rich modeling $p_0 = \la D_1,D_2,f_1,f_2\ra$, where each set
$D_i\subseteq D$ is a set of cards, and $f_i \subseteq D$ is a set of
cards with at most one element.  (In the latter case, not all such
states are reachable: for example, we will have $f_i=c_i$ in all
reachable states.)

The observation of agent $i=1,2$ in a state of one of the above forms is
$O_i(s) = 
 \la \epsilon,c_1, c_2; h_i \ra$. 
Note that we model observability of the card face up through the 
agent's observation of the last action, since agents are assumed 
incapable of observing the private state of agent $0$ directly.
If an agent $i=1,2$ makes a sequence of observations
$\sigma$, then its final observation will have $h_i$ equal to 
its current hand. Thus, we may define the protocol of 
each agent, representing its choice of card at each 
move, by $P_i(\sigma) = \{ \{c\}~|~ c\in h_i\}$ if $h_i$ is not empty, and 
$P_i(\sigma) = \{ \emptyset \} $ otherwise. 

Transitions are given as follows, for states and joint 
external actions as above. For agent $0$, we clearly have 
$\tau_0( \la \epsilon, c_1,c_2 \ra, \epsilon)(1)=1$ in the simple modeling. 
In the rich modeling, 
\[\tau_0( \la \epsilon, c_1,c_2 \ra, \epsilon)( \la
D_1, \ldots D_2, f_1,f_2\ra) = \la D_1 \cup f_1, D_2 \cup f_2, 
c_1, c_2\ra. \] 
For agent $i=1,2$, we take 
\[\tau_0( \la \epsilon, c_1,c_2 \ra, \epsilon)(h_i) = 
h_i \setminus c_i .\] 

The initial states of the system are the states 
\[s= \Big \la \epsilon\cdot \epsilon,~ \emptyset \cdot \epsilon,~ \emptyset\cdot \epsilon; ~ 
   p_0, h_1,h_2 \Big \ra\] where for $i=1,2$ we have $h_1$ and $h_2$
equal to sets of exactly 12 cards.  In the simple modeling, we have
$p_0=1$, and, as noted in Example~\ref{eg:cards}, this set of states
forms a hypercube.  In the rich modeling, we have $p_0 = \la
D\setminus h_1, D\setminus h_2, \emptyset, \emptyset\ra$.  As we noted
previously, this set of states is a full system, but not a hypercube.
\end{example}

In the 
sequel, we will make use of the following observation. 

\begin{lemma} \label{lem:enbld}
Suppose $E$ is a broadcast environment, and $\bP$ is a joint perfect
recall protocol.  Let $r$ and $r'$ be traces of $E$ consistent
with $\bP$ such that $r\sim_i r'$, where $i\in \{0\ldots n\}$.  Then every 
action of agent $i$ that is enabled at $r$ is also enabled at $r'$.
\end{lemma}

The proof is immediate from the definitions. (In the case of agent 0, 
note that $r\sim_0 r'$ implies $O_0(r) = O_0(r')$ 
and use Assumption~\ref{ass:prot}.)

\subsection{Homogeneous broadcast environments}\label{hbs}

Homogeneous broadcast environments are a special case of broadcast
environments.  These environments satisfy the additional,
and quite natural, constraint that agents start in a condition of
ignorance about each others states, and the state of the environment.
Thus, their initial state of knowledge is characterized by a
hypercube system.  We will show that, under the assumption that agents have
perfect recall, their knowledge can also be characterized as a
hypercube system at all subsequent times.

\begin{definition}[Homogeneous broadcast environment] 
A broadcast environment $E$ is {\em homogeneous} if
there exists for each agent $i= 0\ldots n$ a set $I_i\subseteq P_i$ of
{\em initial private states}, such that the set of initial states $I$
of the environment $E$ is the set of all states $\la \epsilon, \ldots,
\epsilon; p_0, \ldots p_n\ra$, where $p_i\in I_i$ for $i=0\ldots n$. 
\end{definition}  
In other words, the set of initial states is isomorphic to the 
hypercube $I_0\times \ldots \times I_n$. 
That is, agents are initially ignorant of each others' states and
the state of the environment.
The environment in Example~\ref{eg:be} is a homogeneous broadcast  
system under the simple modeling (but not under the rich modeling.) 
Of the other examples mentioned in the previous section, battleships and
Stratego satisfy this constraint, but the wise men puzzle, the muddy
children puzzle and Bridge do not. (For example, the initial
configurations of Bridge, i.e. after cards have been dealt but before
bidding, do not form a hypercube because it is not possible for two
players to simultaneously hold the same card.)

We may now introduce the main object of study in this section.

\begin{definition}[Perfect recall homogeneous broadcast frame] 
A \emph{perfect recall homogene\-ous broadcast frame} is any frame
$F_{E,\bP}$ obtained from a joint perfect recall protocol $\bP$ in a
homogeneous broadcast environment $E$.
\end{definition}

We are now in a position to state the main result of this section.

\begin{theorem} \label{thm:hbs}
Every perfect recall homogeneous broadcast frame is isomorphic to a
frame obtained from a disjoint union of systems of the form
$X_0\times X_1\times \ldots \times X_n$.  In particular, every such
frame is 
weakly-directed.
\end{theorem} 

This result establishes a close connection between perfect recall
homogeneous broadcast frames and hypercube systems.  In particular, it
follows that the logic 
\S{} is sound for this class of frames.

For the proof, it is convenient to introduce the following notions.
If 
$r=s_0 s_1 \ldots s_m$ is a trace of a broadcast environment, we
will write $\aseq(r)$ for the sequence 
$\ba(s_0) \ldots \ba(s_m)$ of joint external actions performed in $r$.
If $s= \la a_0,\ldots ,a_n;p_0,\ldots, p_n\ra$ and
$t= \la a_0,\ldots ,a_n;q_0,\ldots, q_n\ra$ are global states with the
same joint external action component, and $i\in \{0\ldots n\}$, 
define $s\join_i t$ to be the state
$\la a_0,\ldots ,a_n;p_0, \ldots,p_{i-1},q_i,p_{i+1},\ldots, p_n\ra$,
that is like $s$ except that agent $i$ has the private state it has in
$t$.  

Note that for $j\neq i$, we have $O_j(s\join_i t) = \la a_0,\ldots
,a_n;p_j\ra=O_j(s)$.  Additionally, $O_i(s\join_i t) = \la a_0,\ldots
,a_n;q_i\ra = O_i(t)$.  More generally, if $r_1=s_0 s_1 \ldots s_m$
and $r_2 = t_0 t_1 \ldots t_m$ are sequences of states of the same
length with $\aseq(r_1) =
\aseq(r_2)$ then we define $r_1\join_i r_2$ to be the sequence $(s_0
\join_i t_0) (s_1 \join_i t_1) \ldots (s_m\join_i t_m)$.  The
following result states a closure condition of the set of traces of a
homogeneous broadcast environment.

\begin{lemma}\label{lem:join}
Let $E$ be a homogeneous broadcast environment, and $\bP$ a joint
perfect recall protocol.  If $r_1$ and $r_2$ are traces in 
$F_{E,\bP}$ with $\aseq(r_1)=\aseq(r_2)$ then 
for any $i$ we have that $r_1\join_i r_2$ is a trace in $F_{E,\bP}$
with $(r_1\join_i r_2) \sim_i r_2$ and for $j \not = i$ we have that
$(r_1\join_i r_2)\sim_j r_1$.
\end{lemma}

\begin{proof}
Note that $\aseq(r) = \aseq(r')$ implies that $r$ and $r'$
have the same length. It is immediate from the comments above that
$(r_1\join_i r_2)\sim_j r_1$ for $j\not = i$ and $(r_1\join_i r_2)
\sim_i r_2$.
It therefore suffices to show that $r_1\join_i r_2$ is a trace. We do
this by induction on the length of the trace $r_1$.

The base case is straightforward. If $r_1$ is a trace of length one,
then it consists of an initial state $s_1$. Similarly, $r_2$ consists
of an initial state $s_2$.  It is immediate from the assumption that
$I= \{\la \epsilon, \ldots, \epsilon\ra \} \times I_0\times \cdots
\times I_n$ that $r_1\join_i r_2 = s_1\join_i s_2$ is a trace.

Assume that the result has been established for traces of length $m$,
and consider traces $r_1$ and $r_2$ of length $m+1$ with 
$\aseq(r_1) = \aseq(r_2)$. 
Write $r_1= r_1' s_1 t_1$ where $t_1$ is the final state of $r_1$ and
$s_1$ is the next-to-final state of $r_1$, and similarly write $r_2=
r_2' s_2 t_2$. 
By the induction hypothesis, $r= r_1' s_1\join_i r_2'
s_2$ is a trace indistinguishable to agent $i$ from $r_2' s_2$, and
indistinguishable to all other agents from $r_1' s_1$.  

Let $\bj_1$ be 
a joint action enabled at $r_1' s_1$ such that $t_1 =
\tau(\bj_1)(s_1)$, and similarly, let $\bj_2$ be 
a joint action enabled at $r_2' s_2$ such that $t_2 =
\tau(\bj_2)(s_2)$.  Note that because state transitions record the
joint external action component of a joint action in the resulting
state, and because $\aseq(r_1) = \aseq(r_2)$, we have $\ba(\bj_1) =
\ba(t_1) = \ba(t_2) = \ba(\bj_2)$.  Write $\la a_0, \ldots, a_n\ra$
for the common joint external action of these states and joint
actions. Then we may also write $\bj_1= \la a_0\cdot b_0,\ldots ,
a_n\cdot b_n\ra$ and $\bj_2= \la a_0\cdot c_0,\ldots , a_n\cdot
c_n\ra$.  To show that $r_1 \join_i r_2$ is a trace we show that the
joint action
\[\bj= \la a_0\cdot
b_0,\ldots ,a_{i-1}\cdot b_{i-1}, a_i\cdot c_i, a_{i+1}\cdot b_{i+1},
\ldots ,a_n\cdot b_n\ra\] is 
enabled at $r$ and satisfies $\tau(\bj)(s_1 \join_i s_2) = t_1\join_i t_2$.

To show that $\bj$ is enabled at $r$ we show that each of its
components is enabled at $r$.  In the case of agents $j \neq i$, we
need to show that the action $a_j\cdot b_j$ of agent $j$ is enabled at
$r$. This follows, using Lemma~\ref{lem:enbld}, from the fact that
$a_j\cdot b_j$ is enabled for agent $j$ at $r_1's_1$, and from the
fact that $r_1's_1 \sim_j r$. For agent $i$, we need to show that the
action $a_i\cdot c_i$ is enabled at $r$. This follows, again using
Lemma~\ref{lem:enbld}, from the fact that $a_i\cdot c_i$ is enabled
for agent $i$ at $r_2's_2$, and from the fact that $r_2's_2 \sim_i r$.

It therefore remains to show that $\tau(\bj)(s_1 \join_i s_2) =
t_1\join_i t_2$. Note first that $\ba(\tau(\bj)(s_1 \join_i s_2)) = 
\ba(\bj) = \ba(\bj_1) = \ba(t_1\join_i t_2)$. Thus, the states
$\tau(\bj)(s_1 \join_i s_2)$ and $t_1\join_i t_2$
record the same joint external action. We show that they also 
have the same private state for each agent. 
In case of agents $j\neq i$, we have 
\[\begin{array}{rcl} 
\bp_j(\tau(\bj)(s_1 \join_i s_2)) & 
= & \tau_j(\ba(\bj),b_j)(\bp_j(s_1 \join_i s_2)) \\ 
& = & \tau_j(\ba(\bj),b_j)(\bp_j(s_1)) \\ 
& = & \bp_j(t_1) \\ 
& = & \bp_j(t_1 \join_i t_2 )
\end{array} 
\] 
In case of agent $i$, we have 
\[\begin{array}{rcl} 
\bp_i(\tau(\bj)(s_1 \join_i s_2)) & 
= & \tau_i(\ba(\bj),c_i)(\bp_i(s_1 \join_i s_2)) \\ 
& = & \tau_i(\ba(\bj),c_i)(\bp_i(s_2)) \\ 
& = & \bp_i(t_2) \\ 
& = & \bp_i(t_1 \join_i t_2 )
\end{array} 
\] 
This completes the proof. 
\end{proof}

Note that because agents observe the most recent joint external
action, if $r$ and $r'$ are traces in 
$F_{\bP,E}$ 
with $r\sim_{i}r'$
then these traces were generated by the same sequence of joint
external actions, i.e., $\aseq(r)=\aseq(r')$.  It follows from this
that if $r$ and $r'$ are in the same connected component of
$F_{\bP,E}$ 
then we also have $\aseq(r)=\aseq(r')$.  In fact, we have
the following stronger result:

\begin{lemma} \label{lem:prod} 
For every trace $r$ we have that:
\begin{itemize}
\item the connected component $F$ of $F_{\bP,E}$ containing
$r$ consists of all traces $r'$ in $F_{\bP,E}$ with $\aseq(r') =
\aseq(r)$ and
\item this connected component is isomorphic to the hypercube
$\Pi_{i=0\ldots n} \{ \{r'\}_i~|~r'\in F\}$.
\end{itemize}
\end{lemma} 

\begin{proof}
1) We prove that $r$ is connected to $r'$ if and only if
   $\aseq(r)=\aseq(r')$. Left to right is immediate from
   Definition~\ref{prframe}. Right to left follows from
   Lemma~\ref{lem:join}. 
 
2) For each agent $i=0\ldots n$, let $r_i$ be any trace of $F_{\bP,E}$
with $\aseq(r_i)=\aseq(r)$.
To show that the connected component is a hypercube, we prove that
  there is an $r'$ such that $r'\sim_i r_i$.  In fact, define $r' =
  (\ldots (r_1 \join_2 r_2) \ldots \join_{n-1} r_{n-1})\join_n r_n$.
  By Lemma~\ref{lem:join}, $r'$ is a trace of $E$, with $r' \sim_i
  r_i$ for all $i=0\ldots n$ and $\aseq(r') = \aseq(r)$.  It is
  immediate that all traces $r'$ of $F_{\bP,E}$ with $\aseq(r') =
  \aseq(r)$ are connected, and that the component containing $r$ is
  isomorphic to the hypercube $\Pi_{i=0\ldots n} \{ \{r'\}_i~|~r'\in
  F\}$.
\end{proof}

This lemma characterizes the sense in which agents' states of knowledge 
at times other than time 0 in a homogeneous broadcast system are 
characterized by a hypercube system. 
Theorem~\ref{thm:hbs} follows immediately  from Lemma~\ref{lem:prod}.

We now obtain a result that provides one final characterization 
of the logic \S. 

\begin{theorem} \label{thm:hbs-wd} 
The logic \S{} is sound and complete for the class of all
homogeneous broadcast frames. 
\end{theorem} 

\begin{proof} 
Soundness is direct from Theorem~\ref{thm:hbs}. For completeness,
suppose that $\phi$ is not a theorem of \S. Since \S{} is
complete for the class of all hypercubes, there exists a hypercube
$H= L_e\times L_1 \times \dots L_n$, where $L_e$ is 
a
singleton, an
interpretation $\pi_H$ on $H$, and a world $w\in S$ such that
$(F(H),\pi_H)\models_w \neg \phi$. 
We show that it is possible to 
construct a homogeneous broadcast environment $E$ whose decomposition into 
a union of Cartesian products contains $H$ as one of its components. 
Indeed $H$ will be the component consisting of all the traces of 
length one, i.e., the component characterizing the initial state 
of knowledge of the agents. 

We define the environment $E = \la S,I,P_0,\tau,O,V \ra$ as
follows. For each agent $i=0\dots n$, 
we take the both the set of external actions $A_i$ and the 
set of internal actions $B_i$ to be the set $\{\epsilon\}$. 
Thus, the set of actions of each agent is also a singleton, 
viz $\{\epsilon \cdot \epsilon\}$. The components of the environment 
are as follows: 
\begin{itemize} 
\item The set of states $S = \la \epsilon, \dots, \epsilon;p_0,\dots,p_n\ra$ 
where $(p_0,\dots,p_n)\in H$. Thus, the set $S_i$ of instantaneous
private states of agent $i$ is exactly the set of local states $L_i$
of agent $i$ in $H$.
\item All states are initial, i.e. $I=S$. 
\item Since the set actions $\act_0$ of agent $0$, the 
environment, is a singleton, the protocol of the environment is the
unique function $P_0: S\rightarrow \act_0$.
\item The transition function $\tau$ is defined by 
$\tau(\bj)(s) = s$ for (the unique) joint action $\bj$ and state $s$. 
(Thus, similarly, the local transition functions $\tau_i$ satisfy 
$\tau_i(\ba,b_i)(p_i) = p_i$ for (the unique) joint external action $\ba$, 
(the unique) internal action $b_i$ and private state $p_i\in L_i$.) 
\item The definition of the observation function $O$ is determined by the fact that $E$ is a broadcast environment, i.e. $O_i( \la 
\epsilon, \dots, \epsilon;p_0,\dots,p_n\ra) = 
\la \epsilon, \dots, \epsilon;p_i\ra$ for each $i=1, \dots, n$. 
\item The valuation $V$ is defined by 
$V( \la 
\epsilon, \dots, \epsilon;p_0,\dots,p_n\ra,q) = \pi_H((p_0,\dots,p_n),q)$.
\end{itemize} 
This is a homogeneous broadcast environment by construction.  It is
now straightforward to establish that for {\em every} joint perfect
recall protocol $\bP$, the connected component of $F_{E,\bP}$
consisting of all traces of length one is isomorphic to
$(F(H),\pi_H)$. (We remark that our choice of action sets and
transition function above are not actually relevant to this conclusion.)
\end{proof} 

One way to understand Theorem~\ref{thm:hbs-wd} is that it states
completeness of \S{} with respect to a class of models, namely
those models obtained by adding an interpretation to a homogeneous
broadcast frame. In these models the interpretation could assign to a
proposition a meaning at a trace that depends not just on the final
state of the trace, but also on prior states and actions.  The proof
of Theorem~\ref{thm:hbs-wd} in fact establishes that \S{} is
complete for a smaller class of models with  underlying homogeneous
broadcast frames, in which the interpretation $\pi$ is derived from the
environment. Given an environment $E$ with valuation $V$, define the
interpretation $\pi_E$ by $\pi_E(r,p) = V(\fin(r),p)$ for traces $r$ of
$E$ and propositions $p\in \atoms$. 

\begin{theorem} 
The logic \S{} is sound and complete for the class of all 
models of the form $(F_{E,\bP},\pi_E)$, where $E$ is a homogeneous 
broadcast environment and $\bP$ is a joint protocol. 
\end{theorem} 

\begin{proof} 
Similar to the proof of Theorem~\ref{thm:hbs-wd}. Note that the construction 
of this proof uses only the initial component of the frame. The valuation of 
the environment may be chosen to operate as required on this initial 
component. 
\end{proof} 

These results are in some respects similar to results of 
Fagin et al.\ \cite{JACM::FaginHV1992,FaginVardi86}. %
They show that that there exist natural classes of systems
with respect to which the logic of knowledge is not characterized by 
S5$_n$, but by a stronger logic $ML_n^-$, that consists of 
S5$_n$    plus the following axiom: 
\[ \beta \And K_i(\beta \implies \neg \alpha) \implies 
  K_1 \neg \alpha \Or \dots \Or K_1 \neg \alpha 
\] 
where $\alpha$ is a \emph{primitive state formula} (intuitively,
describing the assignment associated with the current state, but not
of agents knowledge), and $\beta$ is a \emph{pure knowledge formula}
(intuitively, describing properties of the agents knowledge but not
dealing with the assignment associated with the current state).  We
refer the reader to \cite{JACM::FaginHV1992} for a precise explanation of
these terms.  In particular, one class of systems to which this result
applies is a class of systems in which the assignment is static,
agents communicate by unreliable synchronous message passing and have
perfect recall \cite{FaginVardi86}.

Theorem~\ref{thm:hbs-wd} provides 
another interesting and natural
class of systems that requires additional axioms.
In our result, agents also have perfect recall, but 
the class is otherwise quite different from those 
considered in \cite{JACM::FaginHV1992,FaginVardi86} since our agents 
communicate by reliable broadcast, and we allow the 
assignment to vary significantly from moment to moment. 
The axiom (WD) we need to capture such systems is 
also quite different from that used by Fagin et al.

We remark that it is possible to prove a variant of the results of
this section that deal with full systems rather than hypercubes.  For
this variant, we modify the definition of homogeneity to state that
the initial states of the environment form a full system. Moreover,
instead of Assumption\ref{ass:prot}, we assume that for all states
$s$ and $t$ with the same joint external action, i.e., $\ba(s) =
\ba(t)$, and for all external actions $a_0$ of agent $0$, there exists
an internal action $b_0$ of agent $0$ such that $a_0\cdot b_0 \in
P_0(s)$ iff there exists an internal action $b_0'$ of agent $0$ such
that $a_0\cdot b_0' \in P_0(s)$. (Informally, this means that an
external action of agent $0$ is enabled in $s$ iff it is enabled in
$t$.) The environment and protocol in Example~\ref{eg:be} satisfy both these 
assumptions. 

Under these assumptions, 
Lemma~\ref{lem:join} holds provided we restrict $i$ and $j$ to range
over agents $1$ to $n$ only (i.e., we exclude agent $0$.) The proof is
a trivial adaptation.  Consequently,
we also obtain an analogue of
Lemma~\ref{lem:prod} stating that the connected components of
$F_{E,\bP}$ are full systems.

\section{Conclusions and further work}

In this paper we have formally investigated 
several classes of 
interpreted systems that arise by considering the full Cartesian
product of the local state spaces.  We have argued that these interpreted
systems provide an appropriate model for the initial configurations of
many systems of interest.  Moreover, we have shown that a similar
constraint arises at all later configurations in the special case of
homogeneous broadcast systems. By relating these classes of systems 
to several classes of Kripke frames, we have established that 
a single modal logic, \S, provides a sound and complete 
axiomatization in all these cases. 
On the conceptual level, 
this logic provides a well motivated example 
of interaction among agents'
knowledge. We hope that in the near future \S{} will be just one
of the formal options available in the literature and that, 
with the availability of more examples, a systematic study of types of
interactions can be carried out.

In conducting this work, we have identified 
the interesting class of 
WD equivalence
frames that generates a complete and decidable logic.  The variation
of the canonical model technique we used to prove completeness heavily
relies on the frames being reflexive, symmetric and transitive,
properties guaranteed by the fact that we were analyzing extensions of
S5$_n$.  
This raises the question of whether it is possible to prove similar
results for weaker logics, such as S4$_n$, 
which model agents that
do not have negative introspection capabilities.

Our results leave open many other questions. It should be noted that 
the fact that the same logic \S{} axiomatizes all the 
different classes of semantic structures we have studied is 
is due in part to the limited expressive power  of the language
we have considered. It would be interesting to investigate
more expressive languages containing operators such as 
distributed knowledge and common knowledge \cite{fhmv:rak}. 
In the former case we have already identified
the axiom $\phi \biimplies D_A \phi$ as of interest with respect 
to equivalence 
I frames (Lemma~\ref{da}). 

Although we have shown decidability of \S, the 
precise complexity of this logic remains open.
It would also be of interest to determine which 
of the language extensions contemplated above maintain decidability 
of the logic. 
Finally, for the dynamic model we have considered, extensions of the
language to include temporal operators are of interest. Indeed,
consideration of the logic of knowledge and time in homogeneous
broadcast systems is just one example of a range of unresolved issues
concerning the knowledge of agents 
operating within specific communications models: a great deal
of work remains to be done in the axiomatization of logics of
knowledge with respect to such models.

\bibliography{../bib/biblio}

\end{document}